\documentstyle[latexsym,amssymb,epsfig,aps,floats,eqsecnum,amsfonts]{revtex}

\def\gaq{\raise 0.4ex\hbox{$>$}\kern -0.7em\lower 0.62ex\hbox{$\sim$}}
\def\laq{\raise 0.4ex\hbox{$<$}\kern -0.8em\lower 0.62ex\hbox{$\sim$}}

\def\be{\begin{equation}}
\def\ee{\end{equation}}

\begin{document}

\title{Coalescence of Two Spinning Black Holes: \\ An Effective One-Body Approach}
\author{Thibault Damour}
\address{{\it Institut des Hautes Etudes Scientifiques, 91440 Bures-sur-Yvette, 
France}}
\date{Sept 18, 2001}
\maketitle

\begin{abstract}
We generalize to the case of spinning black holes a recently introduced ``effective 
one-body'' approach to the general relativistic dynamics of binary systems. We show how 
to approximately map the conservative part of the third post-Newtonian (3~PN) dynamics 
of two spinning black holes of masses $m_1$, $m_2$ and spins $\mbox{\boldmath$S$}_1$, 
$\mbox{\boldmath$S$}_2$ onto the dynamics of a non-spinning particle of mass $\mu 
\equiv m_1 \, m_2 / (m_1 + m_2)$ in a certain effective metric $g_{\mu \nu}^{\rm eff} 
(x^{\lambda} ; M , \nu , \mbox{\boldmath$a$})$ which can be viewed either as a 
spin-deformation (with deformation parameter $\mbox{\boldmath$a$} \equiv 
\mbox{\boldmath$S$}_{\rm eff} / M$) of the recently constructed 3~PN effective metric 
$g_{\mu \nu}^{\rm eff} (x^{\lambda} ; M , \nu)$, or as a $\nu$-deformation (with 
comparable-mass deformation parameter $\nu \equiv m_1 \, m_2 / (m_1 + m_2)^2$) of a Kerr 
metric of mass $M \equiv m_1 + m_2$ and (effective) spin $\mbox{\boldmath$S$}_{\rm eff} 
\equiv (1+3 \, m_2 / (4 \, m_1)) \, \mbox{\boldmath$S$}_1 + (1+3 \, m_1 / (4 \, m_2)) \, 
\mbox{\boldmath$S$}_2$. The combination of the effective one-body approach, and of a 
Pad\'e definition of the crucial effective radial functions, is shown to define a 
dynamics with much improved post-Newtonian convergence properties, even for black hole 
separations of the order of $6~GM / c^2$. The complete (conservative) phase-space 
evolution equations of binary spinning black hole systems are written down and their 
exact and approximate first integrals are discussed. This leads to the approximate 
existence of a two-parameter family of ``spherical orbits'' (with constant radius), and 
of a corresponding one-parameter family of ``last stable spherical orbits'' (LSSO). 
These orbits are of special interest for forthcoming LIGO/VIRGO/GEO gravitational wave 
observations. The binding energy and total angular momentum of LSSO's are studied in 
some detail. It is argued that for most (but not all) of the parameter space of two 
spinning holes the approximate (leading-order)
effective one-body approach introduced here
 gives a reliable analytical tool for describing the 
dynamics of the last orbits before coalescence. This tool predicts, in a quantitative 
way, how certain spin orientations increase the binding energy of the LSSO. This leads 
to a detection bias, in LIGO/VIRGO/GEO observations, favouring spinning black hole 
systems, and makes it urgent to complete the conservative effective one-body dynamics 
given here by adding (resummed) radiation reaction effects, and by constructing 
gravitational waveform templates that include spin effects. Finally, our approach
predicts that the spin of the final hole formed by the coalescence of two arbitrarily
spinning holes never approaches extremality.
\end{abstract}

\section{Introduction}\label{sec1}
The most promising candidate sources for the LIGO/VIRGO/GEO/$\ldots$ network of ground 
based gravitational wave (GW) interferometric detectors are coalescing binary systems 
made of massive (stellar) black holes \cite{postnov,FH98,BCT98,PZ99,DIS00}. Signal to 
noise ratio (SNR) estimates \cite{DIS00} suggest that the first detections will 
concern black hole binaries of total mass $\gaq \, 25 M_{\odot}$. Modelling the GW 
signal emitted by such systems poses a difficult theoretical problem because the 
observationally most ``useful'' part of the gravitational waveform is emitted in the 
last $\sim 5$ orbits of the inspiral, and during the ``plunge'' taking place after 
crossing the last stable circular orbit. The transition between (adiabatic) inspiral 
and plunge takes place in a regime where the two bodies are moving at relativistic 
speeds ($v/c \sim 1 / \sqrt 6 \sim 0.4$) and where their gravitational interaction 
becomes (nearly by definition) highly non-linear ($GM / c^2 r \sim 1/6$).

Several authors (notably \cite{cutler93,BCT98}) have taken the view that the 
modelling of this crucial transition between inspiral and plunge is (in the general 
case of comparable-mass systems) beyond the reach of analytical tools and can only be 
tackled by (possibly special-purpose \cite{BCT98}) numerical simulations. By 
contrast, other authors \cite{DIS98,BD99,BD00,DIS00,DJS2,BD01} have introduced new 
``resummation methods'' to improve the analytical description of the last few GW 
cycles near this transition and have argued that these resummed analytical results 
gave a reliable description of the gravitational physics near the transition. The 
purpose of the present paper is to further the latter approach by generalizing the 
(resummed) ``effective one-body'' (EOB)
 methods introduced in \cite{BD99,BD00,DJS2} to the 
case of binary systems of {\em spinning} black holes. Before doing this, we wish to 
clarify what is the rationale for arguing that the ``resummed'' analytical approach 
can describe the last stages of inspiral and the transition between inspiral and 
plunge.

Let us first recall that a lot of effort has been devoted in recent years to the 
analytical computing, by means of post-Newtonian (PN) expansions in powers of $v^2 / 
c^2 \sim GM / c^2 r$, of the equations of motion, and the GW emission, of 
comparable-mass binary systems. The equations of motions have been 
computed to $v^6 / c^6$ ($3$~PN) accuracy by two separate groups, 
\cite{JS98,JS99,DJS1,DJS3} and \cite{BF1,BF2,BF3}, and the two results have been shown 
to agree \cite{DJS4,BF4}. Until recently, there remained
 (in both approaches) an ambiguous 
parameter, $\omega_s$, linked to the problem of regularizing some badly divergent 
integrals arising at the 3~PN level. In a recent work, using an improved
regularization method (dimensional continuation), the first group \cite{DJSd} has
succeeded in determining without ambiguity the value of $\omega_s$, 
namely $\omega_s = 0$. This unique determination of the 3PN equations of motion
is consistent with
an old argument of \cite{D83} showing that it should be possible to model black 
holes by point particles without ambiguity up to the 5~PN level (excluded). 
The emission of GW is unambiguously known to $v^5 / c^5$ 
($2.5$~PN) accuracy \cite{2.5PN}, and has recently been formally computed to
$v^7 / c^7$ ($3.5$~PN) accuracy \cite{3.5PN}, modulo the appearance of several
ambiguous parameters ($\xi, \kappa, \zeta$) linked to the problem of regularizing
some divergences arising at the 3PN level. Dimensional regularization is expected
to determine without ambiguity the values of $\xi, \kappa$, and $ \zeta$, but has
not (yet) been applied to the radiation problem.

We wish to emphasize that such high-order PN results are a {\em necessary}, but {\em 
not} by themselves {\em sufficient}, ingredient for computing with adequate accuracy 
the gravitational waveform of coalescing binaries. Indeed, it was emphasized long ago 
\cite{cutler93} that the PN series (written as straightforward Taylor series in 
powers of some parameter $\varepsilon \sim v/c$) become slowly convergent in the late 
stages of binary inspiral. A first attempt was made in \cite{KWW} to improve the 
convergence of the PN-expanded equations of motion so as to determine the (crucial) 
location of the last stable (circular) orbit (LSO) for comparable-mass systems. 
However, further work \cite{WS93,SW93,DIS98} has shown the unreliability (and 
coordinate-dependence) of this attempt. There is, however, no reason of principle 
preventing the existence of gauge-invariant ``resummation methods'' able to give 
reliable results near the LSO. Indeed, as emphasized in \cite{DIS98} and \cite{DJS2} 
most coordinate-invariant functions (of some invariant quantity $x \sim v^2 / c^2 
\sim GM / c^2 r$) that one wishes to consider when discussing the dynamics and 
GW emission of circular orbits are expected to have a singularity only at the ``light 
ring'' (LR) value of $x$ [last possible {\em unstable} circular orbit]. If we trust 
(for orders of magnitude considerations) the small mass-ratio limit ($\nu \equiv \mu / 
M \equiv m_1 \, m_2 / (m_1 + m_2)^2 \ll 1$), we know that $x_{\rm LR} \simeq 1/3$ is 
smaller by a factor 2 than $x_{\rm LSO} \simeq 1/6$. If the functions $f(x)$ we are 
dealing with are meromorphic functions of $x$, the location of the expected closest 
singularity $(x_{\rm LR})$ determines their radius of convergence. Therefore, we 
expect that, for $x < x_{\rm LR}$, the Taylor expansion of $f(x)$ will converge and will 
behave essentially like ${\sum_n \ (x/x_{\rm LR})^n}$. In particular, one 
expects ${ f (x_{\rm LSO}) \sim \sum_n \ (x_{\rm LSO} / x_{\rm LR})^n 
\sim \sum \ 2^{-n}}$. This heuristic argument suggests a rather slow convergence, but 
the crucial point is to have some convergence, so that the application of suitable 
{\em resummation methods} can be expected to accelerate the convergence and to lead to 
numerically accurate results from the knowledge of only a few terms in the Taylor 
expansion. There exist many types of resummation methods and none of them are of truly 
universal applicability. As a rule, one must know something about the structure of the 
functions $f(x) = f_0 + f_1 \, x + f_2 \, x^2 + \cdots$ one is trying to resum to be 
able to devise an efficient resummation method. Refs.~\cite{DIS98}, \cite{BD99}, 
\cite{BD00} and \cite{DJS2} have studied in detail the various functions that might be 
used to discuss the GW flux and the dynamics of binary systems. This work has led to 
selecting some specific resummation methods, acting on some specific functions. For 
what concerns the GW flux we refer to Fig.~3 of \cite{DIS98} for evidence of the 
acceleration of convergence (near the LSO) provided by a specific resummation method 
combining a redefinition of the {\em GW flux} function with Pad\'e approximants. We 
wish here, for the benefit of the skeptics, to exhibit some of the evidence for the 
acceleration of convergence (near the LSO) in the description of the 2-body {\em 
dynamics} provided, at the 3~PN level, by a resummation method defined by combining 
\cite{BD99} and \cite{DJS2}. Specifically, we mean the combination of the 
effective-one-body (EOB) approach (further discussed below) and of a suitable Pad\'e 
resummation of the effective radial potential at the $n$~PN level: $A_{P_n} (u) = 
P_n^1 \, [T_{n+1} (A(u))]$ (see below).

 Let us consider a sequence of {\em circular} 
orbits, near the LSO and for two non-spinning black holes. In the EOB approach the
circular orbits are obtained by minimizing a certain effective radial potential, defined
by fixing the total orbital angular momentum $L$ in the Hamiltonian. The most natural
variable defining the one-parameter sequence of circular orbits is then simply the
angular momentum $L$. It is therefore natural, for the purpose of this work,
 to measure the separation 
between the two holes (in a gauge-invariant and approximation-independent) way by 
{\em conventionally} defining a $\ell$-radius $R_{\ell} \equiv GM \, r_{\ell}$, 
such that the (invariantly defined) total orbital angular momentum $L \equiv G \mu M 
\ell$ is given by $\ell^2 \equiv r_{\ell}^2 / (r_{\ell} - 3)$, i.e. by the relation 
holding for a test particle in a Schwarzschild spacetime. [Here, and in the following, we 
shall often set $c=1$ and/or $G=1$, except in some (final) formulas where it might be 
illuminating to reestablish the dependence on $c$ and/or $G$.] As the problem is to know 
whether the resummation 
method of the PN-expanded two-body dynamics is efficient, we compare in Table~\ref{Tab1} 
the 
total energies ${\cal E}_{\rm real}$ of the binary system, computed using 1~PN, 2~PN 
and 3~PN information, for circular orbits at $\ell$-radii $r_{\ell} = 12, 11, 10, 9, 
8, 7$, and 6. We give the values of the binding energy per unit (total) mass, $e 
\equiv ({\cal E} / M) - 1$, for the equal-mass case ($m_1 = m_2$; $\nu = 1/4$). 

\begin{table}
\begin{center}
\begin{tabular}{cccccccc}
$r_{\ell}$  & $12$ & $11$ & $10$ & $9$ & $8$ & $7$ & $6$ \\ 
\hline
$\ell$ & $4.$ & $3.889$ & $3.780$ & $3.674$ & $3.578$ & $3.5$ & $3.464$ \\
\hline
$100 \, e_{1 \, {\rm PN}}$ & $-0.9482$ & $-1.020$ & $-1.101$ & $-1.193$ & $-1.291$ & 
$-1.387$ & $-1.440$ \\
\hline
$100 \, e_{2 \, {\rm PN}}$ & $-0.9441$ & $-1.015$ & $-1.094$ & $-1.183$ & $-1.277$ & 
$-1.366$ & $-1.412$ \\
\hline
$100 \, e_{3 \, {\rm PN}}^{(\omega_s = 0)}$ & $-0.9412$ & $-1.011$ & $-1.088$ & 
$-1.174$ & $-1.264$ & $-1.346$ & $-1.388$
\end{tabular}
\medskip
\caption{Binding energies $e \equiv ({\cal E}_{\rm real} / M) - 1$ for circular orbits of 
equal-mass (non-spinning) binary systems near the LSO. The invariant dimensionless 
$\ell$-radius $r_{\ell} \equiv R_{\ell} / GM$ is defined in the text. The binding energy 
is 
computed from the Pad\'e-resummed Effective-one-body Hamiltonian at three successive 
post-Newtonian approximations: 1~PN, 2~PN and 3~PN (with $\omega_s = 0$ ).}
\label{Tab1}
\end{center}
\end{table}

The numbers displayed in Table~\ref{Tab1} illustrate the efficiency of the resummation 
method advocated in \cite{BD99,DJS2}. For $r_{\ell} = 12$ the fractional difference in 
binding energy between the 1~PN approximation and the 3~PN one is 0.74\%, while even 
for $r_{\ell} = 6$ this difference is only 3.6\%. These numbers indicate that, even 
near the LSO, the Pad\'e-improved effective-one-body approach is a rationally sound 
way of computing the 2-body dynamics. There are no signs of numerical unreliability, 
as there were in the calculations based on the straightforward coordinate-dependent, 
PN-expanded versions of the equations of motion \cite{KWW} or of the Hamiltonian 
\cite{WS93,SW93} (which gave results differing by ${\cal O}$ (100\%) among themselves, 
and as one changed the PN order). We shall see below that the robustness of the 
PN-resummation exhibited in Table~\ref{Tab1} extends to a large domain of the parameter 
space of spinning black holes.

As we do not know the exact result, and as current numerical simulations do not give 
reliable information about the late stages of the quasi-circular orbital dynamics of 
two black holes (see below), the kind of internal consistency check exhibited in 
Table~\ref{Tab1} 
is about the only evidence we can set forth at present. [Note that, from a logical 
point of view, the situation here is the same as for numerical simulations: in absence 
of an exact solution (and of experimental data) one can only do internal convergence 
tests.] Ideally, it would be important to extend the checks of Table~\ref{Tab1} to the 
4~PN level (to confirm the trend and see a real sign of convergence to a limit) but 
this seems to be an hopelessly difficult task with present analytical means.
 Finally, let us note that the fact that one can concoct 
many ``bad'' ways of using the PN-expanded information near the LSO (exhibiting as 
badly divergent results as wished) is not a valid argument against the reliability of 
the specific resummation technique used in \cite{DIS98,BD99,DJS2} and here. An 
ambiguity problem would arise only if one could construct two different resummation 
methods, both exhibiting an internal ``convergence'' (as the PN order increases) as 
good as that illustrated in Table~\ref{Tab1}, but yielding very different predictions 
for physical observables near the LSO. This is not the case at present because the 
comparative study of \cite{DJS2} (see Table~I there) has shown that the EOB 
approach exhibited (when $\omega_s \ne -9$) significantly better PN 
convergence than a panel of other invariant resummation methods.

In the following we take for granted the soundness of the effective-one-body 
resummation approach and we show how to generalize it to the case of two 
(moderately) spinning 
black holes. Let us first recall that the basic idea of the EOB 
approach was first developed in the context of the electromagnetically interacting 
quantum two-body problem \cite{BIZ}, \cite{T1} (see also \cite{B00}). A first attempt 
to deal with the gravitationally interacting two-body problem (at the 1~PN level) was 
made in \cite{T2} (see also \cite{FT01}). A renewed EOB approach (which 
significantly differs from the general framework set up by Todorov and coworkers 
\cite{T1,T2}) was introduced in \cite{BD99}. The latter reference showed how to apply 
this method at the 2~PN level. It was then used to study the transition between the 
inspiral and the plunge for comparable masses, and, in particular, to construct a complete 
waveform covering 
the inspiral, the plunge and the final merger \cite{BD00} (see \cite{DIS3} for the 
physical consequences of this waveform). More recently, Ref.~\cite{DJS2} showed how to 
extend the EOB approach to the 3~PN level (this required a non trivial 
generalization of the basic idea).

Before entering the details of our way of introducing the spin degrees of freedom
in the EOB approach, let us state our general view of the usefulness of the EOB
method in this context. As we shall discuss below, the present work (which incorporates
spin effects at leading PN order)
 can only be expected to give physically reliable results in the case
of moderate spins ( $\widehat{a}_p \alt 0.3$, see below). However, the EOB approach,
far from being a rigid structure, is extremely flexible. One can modify the basic
functions (such as $A(u)$) determining the EOB dynamics by introducing new
parameters corresponding to (yet) uncalculated higher PN effects.
 [ These terms become important only for orbits closer than $ 6 \, G M$, and/or  for
 fast-spinning holes.] Therefore, when either higher-accuracy analytical
 calculations are performed or numerical relativity
becomes able to give physically relevant data about the interaction of 
(fast-spinning) black holes, we expect that it will be possible to complete the
current EOB Hamiltonian so as to incorporate this information. As the parameter space
of two  spinning black holes (with arbitrarily oriented spins) is very large,
numerical relativity will never be able, by itself, to cover it densely. We think,
however, that a suitable ``numerically fitted'' (and, if possible, 
``analytically extended'') EOB Hamiltonian should be able to fit the needs of
upcoming GW detectors. The present work should be viewed as a first step in
this direction.

The present paper is organized as follows. In Section~\ref{sec2} we show how to 
incorporate (in some approximation) the spin degrees of freedom of each black hole 
within a 3~PN-level, resummed effective one-body approach. In Section~\ref{sec3} we 
study some of the predictions of our resummed dynamics, notably for what concerns the 
location of the transition between the inspiral and the plunge. Section~\ref{sec4} 
contains our concluding remarks.


\section{Effective one-body approach, effective spin and a deformed Kerr 
metric}\label{sec2}

\subsection{Effective one-body approach}\label{ssec2.1}

Let us recall the basic set up of the effective one-body (EOB) approach. One starts from the 
(PN-expanded) two-body equations of motion, which depend on the dynamical variables of 
two particles. One separates the equations of motion in a ``conservative part'', and a 
``radiation reaction part''. Though this separation is not well-defined at the exact 
(general relativistic) level it is not ambiguous at the 3~PN level (in the 
conservative part) which we shall consider here \footnote{We expect real ambiguities 
to arise only at the $v^{10} / c^{10} \sim$ 5~PN level, because this corresponds to 
the {\em square} of the leading, $v^5 / c^5 \sim$ 2.5~PN, radiation reaction terms.}. 
We shall henceforth consider only the conservative part of the dynamics. [We leave to 
future work the generalization to spinning black holes of the definition of {\em 
resummed} radiation reaction effects which was 
achieved in \cite{BD00} for non-spinning black holes.] It has been explicitly shown 
that the 3~PN\footnote{Henceforth, `3~PN' will mean the conservative 3~PN dynamics, 
i.e. $N + 1 \, {\rm PN} + 2 \, {\rm PN} + 3 \, {\rm PN}$.} dynamics is Poincar\'e 
invariant 
\cite{DJS3}, \cite{BF2}. The ten first integrals associated to the ten generators of 
the Poincar\'e group were constructed in \cite{DJS3} (see also \cite{BF4}). In 
particular, we have the ``center of mass'' vectorial constant $\mbox{\boldmath$K$} = 
\mbox{\boldmath$G$} - t \, \mbox{\boldmath$P$}$. This constant allows one to define 
the center of mass frame, in which $\mbox{\boldmath$K$} = 0$, which implies 
$\mbox{\boldmath$P$} = 0$ and $\mbox{\boldmath$G$} = 0$. We can then reduce the 
PN-expanded two-body dynamics to a PN-expanded one-body dynamics by considering the 
relative motion in the center of mass frame. This reduction leads to a great 
simplification of the dynamics.

Indeed, the full 3~PN Hamiltonian in an arbitrary reference frame \cite{DJS3} contains 
${\cal O} (100)$ terms, while its center-of-mass-reduced version contains only 24 terms. 
However, 
this simplification is, by itself, insufficient for helping in any way the crucial 
problem of the slow convergence of the PN expansion. One should also mention that the 
use of an Hamiltonian framework (like the ADM formalism used in 
\cite{JS98,JS99,DJS2,DJS3,DJSd}) is extremely convenient (much more so than an approach 
based on the harmonic-coordinates equations of motion, as in \cite{BF1,BF2,BF3,BF4}). 
Indeed, on the one hand it simplifies very much the reduction to the center-of-mass 
relative dynamics (which is trivially obtained by setting $\mbox{\boldmath$p$}_{\rm 
rel} = \mbox{\boldmath$p$}_1 = - \mbox{\boldmath$p$}_2$), and on the other hand it 
yields directly (without guesswork) an action principle for the dynamics\footnote{Note 
that a subtlety arises at 3~PN \cite{JS98} in that the Hamilton action principle 
involves {\em derivatives} of the phase space variables. However, it was shown in 
\cite{DJS1} how to reduce the problem to an ordinary Hamiltonian dynamics by means of 
a suitable ${\cal O} (v^6 / c^6)$ shift of phase-space variables. We henceforth assume 
that we work with the shifted variables defined in \cite{DJS1}.}. We shall find also 
below that an Hamiltonian approach is very convenient for dealing with the spin 
degrees of freedom.

Up to now, we only mentionned the dynamics of the orbital degrees of freedom, i.e. (in 
the order-reduced Hamiltonian formalism) the (ADM-coordinate) positions and momenta 
$\mbox{\boldmath$x$}_1 , \mbox{\boldmath$x$}_2 , \mbox{\boldmath$p$}_1 , 
\mbox{\boldmath$p$}_2$ of the two black holes\footnote{We recall that the high-order 
perturbative, PN-expanded, calculations of the dynamics of two non-spinning compact 
objects model these objects by delta-function (monopole) sources. The supports of 
these delta-functions define the coordinate ``positions'' of the compact objects. As 
explained in \cite{D83} these ``positions'' physically correspond to some ``centers of 
the gravitational field'' generated by the objects.}. After reduction to the 
center-of-mass frame ($\mbox{\boldmath$P$} = \mbox{\boldmath$p$}_1 + 
\mbox{\boldmath$p$}_2 = 0$), and to the relative dynamics ($\mbox{\boldmath$x$} \equiv 
\mbox{\boldmath$x$}_1 - \mbox{\boldmath$x$}_2$, $\mbox{\boldmath$p$} \equiv  
\mbox{\boldmath$p$}_1 = - \mbox{\boldmath$p$}_2$), one ends up with a canonical pair 
$\mbox{\boldmath$x$}$, $\mbox{\boldmath$p$}$ of phase-space variables.

The addition of spin degrees of freedom on each black hole is, a priori, a rather 
complicated matter. If one wished to have a {\em relativistically covariant} 
description of the dynamics of two spinning objects, one would need not only to add, 
in Einstein's equations, extra (covariant) source terms proportional to suitable 
derivatives of delta-functions (spin dipoles), but also to enlarge the two-body action 
principle to incorporate the spin variables. The first task is doable, and has been 
performed (to the lowest order) in several works, such as 
Refs.~\cite{tulczyjew,infeld-plebanski,D75,D82,martin}. For other works on the 
relativistic equations of motion of black holes or extended bodies (endowed with spin and 
higher multipole moments) see \cite{D'Eath75,thorne-hartle,DSX}. Recently, 
Ref.~\cite{Tagoshi} has tackled the next-to-leading order contribution to 
spin-orbit effects.
[We consider here only 
the case of interacting, comparable massive objects. The problem of a spinning {\em 
test} particle in an external field is simpler and has been dealt with by many 
authors, such as Mathisson \cite{mathisson}, Papapetrou \cite{papapetrou}, etc$\ldots$] On 
the other hand, the second 
task (incorporating the spin degrees of freedom in an action principle) is quite 
intricate. First, it has been found that, within a relativistically covariant set up 
for a spinning particle, the Lagrangian describing the {\em orbital} motion could not, 
even at lowest order in the spin, be taken as an ordinary Lagrangian $L (x,\dot x)$, 
but needed to be a higher-order one $L (x,\dot x , \ddot x , \ldots)$ \cite{D82}. 
Second, a relativistically covariant treatment of the {\em spin} degrees of freedom is 
an intricate matter, involving all the subtleties of constrained dynamical systems, 
even in the simplest case of a free relativistic top \cite{regge}.
Contrary to the case of the spin-independent EOB where it was easy to use the test-mass
results to constrain the EOB Hamiltonian, the Mathisson-Papapetrou dynamics of 
spinning test masses in external gravitational fields is rather complicated and
cannot easily be used to constrain the spin-dependent EOB Hamiltonian.
(It might, however, be interesting to try to do so.)
 Fortunately, there 
is a technically much lighter approach which bypasses these problems and simplifies 
{\em both} the description of orbital degrees of freedom and that of spin degrees of 
freedom. This approach is {\em not} manifestly relativistically covariant. This lack 
of {\em manifest} Poincar\'e covariance is not (in principle)
 a problem at all for two reasons: (i) 
it does not prevent the expected global Poincar\'e covariance of the two-body dynamics 
to be realized as a phase-space symmetry (as was explicitly proven, at 3~PN, for the 
orbital degrees of motion in Ref.~\cite{DJS3}), and (ii) as we are, at this stage, 
mainly interested in the description of the relative motion in a specific 
(center-of-mass) Lorentz frame, there is no physical need to enforce any boost 
invariance. This non covariant approach to the gravitational interaction of spinning 
objects stems from the classic work of Breit on the electromagnetic interaction of 
(quantum) spinning electrons (see, e.g. \cite{LL}) and has been developed in a 
sequence of papers \cite{BGH66,BOC70,BOC75,BOC79}. [The change of variables needed to 
pass from the covariant, higher-order Lagrangian description to the non-covariant, 
ordinary Lagrangian has been discussed in several papers, e.g. 
\cite{regge,BOC74GRG,DS88}.] In the present paper, we shall combine the non 
manifestely covariant, ADM-Hamiltonian treatment of the orbital degrees of freedom of 
\cite{JS98,DJS3}, with the similarly non covariant, but Hamiltonian, treatment of the 
spin degrees of freedom of \cite{BGH66,BOC75}. Moreover, as is explained below, we 
shall improve upon \cite{BOC70,BOC75,BOC79} in using a direct Poisson-bracket 
treatment of the dynamical spin variables.

Finally, our starting point (for the effective one-body approach) is a {\em 
PN-expanded} Hamiltonian for the {\em relative} motion of two spinning objects of the 
form
\begin{eqnarray}
\label{eq2.1}
H_{\rm real}^{\rm PN} (\mbox{\boldmath$x$} , \mbox{\boldmath$p$} , 
\mbox{\boldmath$S$}_1 , \mbox{\boldmath$S$}_2) &= &H_{\rm orb}^{\rm PN} 
(\mbox{\boldmath$x$} , \mbox{\boldmath$p$}) + H_S^{\rm PN} (\mbox{\boldmath$x$} , 
\mbox{\boldmath$p$} , \mbox{\boldmath$S$}_1 , \mbox{\boldmath$S$}_2) \nonumber \\
&+ &H_{SS}^{\rm PN} (\mbox{\boldmath$x$} , \mbox{\boldmath$p$} , \mbox{\boldmath$S$}_1 
, \mbox{\boldmath$S$}_2) + H_{SSS}^{\rm PN} (\mbox{\boldmath$x$} , \mbox{\boldmath$p$} 
, \mbox{\boldmath$S$}_1 , \mbox{\boldmath$S$}_2) + \cdots 
\end{eqnarray}

Here, $H_{\rm orb}^{\rm PN}$ denotes the PN-expanded {\em orbital} Hamiltonian, which 
is the sum of the free Hamiltonian $H_0 = \sqrt{m_1^2 \, c^4 + \mbox{\boldmath$p$}_1^2 
\, c^2} + \sqrt{m_2^2 \, c^4 + \mbox{\boldmath$p$}_2^2 \, c^2}$ and of the  
monopolar interaction Hamiltonian $H_{\rm int}^m$ generated by the source terms 
proportional to the masses. Before the reduction to the center-of-mass frame $H_{\rm 
int}^m$ has the symbolic structure: $H_{\rm int}^m \sim m_1 \, m_2 + m_1^2 \, m_2 + 
m_1 \, m_2^2 +  m_1^3 \, m_2 + m_1^2 \, m_2^2 + m_1 \, m_2^3 + \cdots \sim m_1 \, m_2 
(1 + m_1 + m_2 + (m_1 + m_2)^2 + (m_1 + m_2)^3 + \cdots)$. It is explicitly known up 
to the 3~PN level (i.e. up to velocity-independent terms $\propto \, m_1 \, m_2 (m_1 + 
m_2)^3$). After reduction to the center-of-mass frame the PN expansion of $H_{\rm 
orb}^{\rm PN}$ reads (with $M \equiv m_1 + m_2$, $\mu \equiv m_1 \, m_2 / M$, $\nu 
\equiv \mu / M \equiv m_1 \, m_2 / (m_1 + m_2)^2$, $r \equiv \vert \mbox{\boldmath$x$} 
\vert$, $\widehat{\mbox{\boldmath$p$}} \equiv \mbox{\boldmath$p$} / \mu$, 
$\widehat{\mbox{\boldmath$x$}} \equiv \mbox{\boldmath$x$} / GM$)
\be
\label{eq2.2}
H_{\rm orb}^{\rm PN} (\mbox{\boldmath$x$} , \mbox{\boldmath$p$}) = M c^2 + H_N 
(\mbox{\boldmath$x$} , \mbox{\boldmath$p$}) + H_{1 \, {\rm PN}} (\mbox{\boldmath$x$} , 
\mbox{\boldmath$p$}) + H_{2 \, {\rm PN}} (\mbox{\boldmath$x$} , \mbox{\boldmath$p$}) + 
H_{3 \, {\rm PN}} (\mbox{\boldmath$x$} , \mbox{\boldmath$p$}) \, ,
\ee
\be
\label{eq2.3}
H_N (\mbox{\boldmath$x$} , \mbox{\boldmath$p$}) = \frac{\mbox{\boldmath$p$}^2}{2 \mu} 
- \frac{GM \, \mu}{r} = \mu \left[ \frac{1}{2} \, \widehat{\mbox{\boldmath$p$}}^2 - 
\frac{1}{\widehat r} \right] \, ,
\ee
\be
\label{eq2.4}
H_{1 \, {\rm PN}} (\mbox{\boldmath$x$} , \mbox{\boldmath$p$}) = \frac{\mu}{c^2} \left[ 
\frac{1}{8} \, (3 \nu - 1) \, \widehat{\mbox{\boldmath$p$}}^4 - \frac{1}{2} \left[ (3+\nu) 
\, \widehat{\mbox{\boldmath$p$}}^2 + \nu (\mbox{\boldmath$n$} \cdot 
\widehat{\mbox{\boldmath$p$}})^2 \right] \, \frac{1}{\widehat r} + \frac{1}{2 \, 
\widehat{r}^2} \right] \, ,
\ee
\be
\label{eq2.5}
H_{2 \, {\rm PN}} (\mbox{\boldmath$x$} , \mbox{\boldmath$p$}) = \frac{\mu}{c^4} \left[ 
\frac{1}{16} \, (1 - 5 \nu + 5 \nu^2) \, \widehat{\mbox{\boldmath$p$}}^6 + \cdots - 
\frac{1}{4} \, (1+3\nu) \, \frac{1}{\widehat{r}^3} \right] \, ,
\ee
\begin{eqnarray}
\label{eq2.6}
H_{3 \, {\rm PN}} (\mbox{\boldmath$x$} , \mbox{\boldmath$p$}) &= &\frac{\mu}{c^6} 
\Biggl[ \frac{1}{128} \, (-5 + 35 \nu - 70 \nu^2 + 35 \nu^3) \, 
\widehat{\mbox{\boldmath$p$}}^8 + \cdots \nonumber \\
&+ &\left[ \frac{1}{8} + \left( \frac{109}{12} - \frac{21}{32} \, \pi^2 + 
\omega_s \right) \nu \right] \, \frac{1}{\widehat{r}^4} \Biggl] 
\, .
\end{eqnarray}

We have exhibited (for illustration) in Eqs.~(\ref{eq2.5}) and (\ref{eq2.6}) only the 
first and the last terms. We refer to \cite{DS88} for the full (center-of-mass) 2~PN 
Hamiltonian (7 terms in all), and to \cite{JS98,DJS1,DJS3} for the full 
(center-of-mass) 3~PN Hamiltonian (11 terms in all). Our effective one-body treatment 
will take into account the {\em full} 2~PN and 3~PN structures, but in a very 
streamlined way which will be explicitly displayed below.

The other terms in Eq.~(\ref{eq2.1}) denote the various spin-dependent contributions 
to the Hamiltonian: respectively the terms linear $(H_S^{\rm PN})$, quadratic 
$(H_{SS}^{\rm PN})$, cubic $(H_{SSS}^{\rm PN})$, etc$\ldots$ in the spins 
$\mbox{\boldmath$S$}_1$, $\mbox{\boldmath$S$}_2$. Before reduction to the 
center-of-mass frame they have the symbolic structure:
\begin{eqnarray}
\label{eq2.7}
H_S^{\rm PN} &\sim &S_1 \, m_2 (1 + m_1 + m_2 + ( m_1 + m_2)^2 + \cdots) \nonumber \\
&+ &S_2 \, m_1 (1 + m_1 + m_2 + ( m_1 + m_2)^2 + \cdots) \, ,
\end{eqnarray}
\begin{eqnarray}
\label{eq2.8}
H_{SS}^{\rm PN} &\sim &S_1^2 \, m_2 \left( \frac{1}{m_1} + 1 + m_1 + m_2 + \cdots 
\right) + S_1 \, S_2 (1 + m_1 + m_2 + \cdots) \nonumber \\
&+ &S_2^2 \, m_1 \left(\frac{1}{m_2} + 1 + m_1 + m_2 + \cdots \right) \, ,
\end{eqnarray}
etc$\ldots$ We shall explain below the occurence of the terms quadratic in the spins 
and inversely proportional to a mass. In contradistinction with the case of the 
orbital Hamiltonian which has been worked out with a high PN accuracy, only the simplest 
spin-dependent terms have been explicitly derived, namely the lowest PN-order term in 
$H_S^{\rm PN}$, whose center-of-mass reduction reads \cite{BOC75}
\be
\label{eq2.9}
H_S^{\rm PN} (\mbox{\boldmath$x$}, \mbox{\boldmath$p$}, \mbox{\boldmath$S$}_1 , 
\mbox{\boldmath$S$}_2) = \frac{2 G}{c^2 \, r^3} \left[ \left( 1 + \frac{3}{4} \, 
\frac{m_2}{m_1} \right) \mbox{\boldmath$S$}_1 + \left( 1 + \frac{3}{4} \, 
\frac{m_1}{m_2} \right) \mbox{\boldmath$S$}_2 \right] \cdot (\mbox{\boldmath$x$} 
\times \mbox{\boldmath$p$}) + {\cal O} \left( \frac{1}{c^4} \right) \, ,
\ee
and the lowest PN-order one-graviton exchange contribution to the bilinear term 
$(\propto \, S_1 \, S_2)$ in $H_{SS}^{\rm PN}$. Ref.~\cite{Tagoshi} contains some
information about the ${\cal O} ( \frac{1}{c^4})$ corrections in
Eq.~(\ref{eq2.9}), but, because of the use of different gauge and spin conditions,
not in a form which can be directly used to derive these corrections.
We shall discuss the spin-bilinear contribution $(\propto \, S_1 \, S_2)$ below,
 together 
 with the leading spin-quadratic contributions $\propto \, S_1^2 \, m_2 / m_1 + S_2^2 
\, m_1 / m_2$.

Before going further, let us make clear that, before and after any type of resummation, 
the dynamics entailed by the Hamiltonians we shall consider $H (\mbox{\boldmath$x$}, 
\mbox{\boldmath$p$}, \mbox{\boldmath$S$}_1 , \mbox{\boldmath$S$}_2)$ follow, for all 
degrees of freedom, from the basic Poisson brackets
\be
\label{eq2.10}
\{ x^i , p_j \} = \delta_j^i \, ,
\ee
\be
\label{eq2.11}
\{ S_1^i , S_1^j \} = \varepsilon^{ijk} \, S_1^k \, ,
\ee
\be
\label{eq2.12}
\{ S_2^i , S_2^j \} = \varepsilon^{ijk} \, S_2^k \, ,
\ee
\be
\label{eq2.13}
0 = \{ x^i , x^j \} = \{ p_i , p_j \} = \{ S_1^i , S_2^j \} = \{ x^i , S_1^j \} = \{ 
x^i , S_2^j \} = \{ p_i , S_1^j \} = \{p_i , S_2^j \} \, .
\ee
The (real) time evolution of any dynamical quantity $f (\mbox{\boldmath$x$}, 
\mbox{\boldmath$p$}, \mbox{\boldmath$S$}_1 , \mbox{\boldmath$S$}_2)$ is given by
\be
\label{eq2.14}
\frac{d}{dt} \, f (\mbox{\boldmath$x$}, \mbox{\boldmath$p$}, \mbox{\boldmath$S$}_1 , 
\mbox{\boldmath$S$}_2) = \{ f , H_{\rm real} \} \, ,
\ee
where the Poisson bracket (PB) $\{ f , H_{\rm real} \}$ is computed from the basic 
PB's (\ref{eq2.10})--(\ref{eq2.13}) by using the standard PB properties (skew 
symmetry: $\{ f,g \} = - \{ g,f \}$, Leibniz rule: $\{ f,gh \} = \{ f,g \} \, h + g \, 
\{ f,h \}$, and the Jacobi identity: $\{ f , \{ g,h \}\} + \{ g , \{ h,f \}\} + \{ h , 
\{ f,g \}\} = 0$). The simplest way to prove the statements 
(\ref{eq2.10})--(\ref{eq2.14}) is to consider our dynamics as the classical limit of 
the quantum dynamics of a system of gravitationally interacting spinning particles. 
Surprisingly, though Refs.~\cite{BGH66,BOC70} derived (\`a la Breit) the spin-dependent 
contributions to the Hamiltonian by a quantum route, they never noticed that 
they could very simply derive the spin evolution equations by using the PB's 
(\ref{eq2.11}), (\ref{eq2.12}). They had to go back to a Lagrangian formalism and add 
some explicit spin kinetic energy terms $\left( \frac{1}{2} \, I_1 \, 
\mbox{\boldmath$\omega$}_1^2 + \frac{1}{2} \, I_2 \, \mbox{\boldmath$\omega$}_2^2 
\right)$ to derive the spin evolution equations. Note also that we have kept the label 
``real'' on the Hamiltonian in Eq.~(\ref{eq2.14}) to distinguish the evolution with 
respect to the real time (associated with the original two-body system) from the 
evolution generated by the effective Hamiltonian to be introduced below (which is 
associated with an auxiliary, effective time).

Before generalizing it, by including the spin degrees of freedom, let us recall the 
results of \cite{BD99} (2~PN level) and \cite{DJS2} (3~PN level) concerning the 
effective one-body ``upgrading'' of the PN-expanded orbital Hamiltonian $H_{\rm 
orb}^{\rm PN} (\mbox{\boldmath$x$} , \mbox{\boldmath$p$})$. Again, the simplest way to 
motivate it is to think of our dynamics as the classical limit of a quantum dynamics 
defined by some hermitian Hamiltonian operator $H_{\rm orb} (\mbox{\boldmath$x$} , - i 
\hbar \mbox{\boldmath$\nabla$})$. We are mainly interested in the bound states of 
$H_{\rm orb}$. It is crucial to note that the orbital Hamiltonian 
(\ref{eq2.2})--(\ref{eq2.6}) is symmetric under an $O(3)$ group (corresponding to 
arbitrary rotations of the relative position $\mbox{\boldmath$x$} = 
\mbox{\boldmath$x$}_1 - \mbox{\boldmath$x$}_2$, in the center-of-mass frame). 
Therefore the quantum (and classical) bound states will be labelled (besides parity) 
by only two quantum numbers: (i) the total orbital angular momentum 
$\mbox{\boldmath$L$}^2 = L (L + \hbar)$, and (ii) some ``principal quantum number'' 
$N$ (such that $(N-L) / \hbar$ counts the number of nodes in the radial relative wave 
function). Both $L$ and $N$ are quantized in units of $\hbar$. The full list of 
two-body bound states is thereby encoded in the formula giving the bound state energy 
as a function of the two quantum numbers $L$ and $N : E_{\rm real} = E_{\rm real} 
(L,N) = M c^2 - \frac{1}{2} \, \mu (G \,m_1 \, m_2)^2 / N^2 + E_{1 \, {\rm PN}} (L,N) 
+ E_{2 \, {\rm PN}} (L,N) + E_{3 \, {\rm PN}} (L,N)$. The basic idea of the effective 
one-body method is to map (in a one-to-one manner) the discrete set of real two-body 
bound states $E_{\rm real} (L,N)$ onto the discrete set of bound states of an 
auxiliary (``effective'') one-body Hamiltonian $H_{\rm eff} (\mbox{\boldmath$x$}_{\rm 
eff} , \mbox{\boldmath$p$}_{\rm eff}$). Because of the special labelling by (only) two 
integer quantum numbers $L/\hbar$, $N/\hbar$ one is naturally led to imposing that: 
(i) the EOB Hamiltonian be spherically symmetric\footnote{We are making 
this very explicit because some people, when they hear about the EOB 
approach, think that the effective metric describing the one-body dynamics should, at 
some level of approximation, include some Kerr-like features to model the 
velocity-dependent two-body interactions. This is not true for the orbital dynamics, 
whatever be the PN accuracy level. On the other hand, we shall see that we need 
Kerr-like features to accomodate the intrinsic spin effects.}, and (ii) the integer 
valued quantum numbers be identified in the two problems, i.e. $L/\hbar = L_{\rm eff} 
/ \hbar$ and $N/\hbar = N_{\rm eff} / \hbar$. On the other hand, one can (and one a 
priori should) leave free a (one-to-one) continuous function $f$ mapping the real 
energies onto the effective ones: $E_{\rm eff} (L,N) = f (E_{\rm real} (L,N))$. 
Evidently, for this method to buy us anything we wish the effective dynamics to be 
significantly simpler than the original $H_{\rm real} (\mbox{\boldmath$x$} , 
\mbox{\boldmath$p$})$, and, in particular, to reduce, in some approximation, to the 
paradigm of the simplest gravitational one-body problem, namely the dynamics of an 
(effective) test particle in some (to be determined) effective metric $g_{\mu 
\nu}^{\rm eff} (x_{\rm eff}^{\lambda})$. Remarkably enough, it was found in 
\cite{BD99} that such a mapping between the very complicated real two-body orbital 
2~PN Hamiltonian (\ref{eq2.2})--(\ref{eq2.5}) and the usual (``geodesic'') dynamics of 
a test particle of mass $\mu = m_1 \, m_2 / (m_1 + m_2)$ in a very simple (spherically 
symmetric) effective metric
\be
\label{eq2.15}
ds_{\rm eff}^2 = - A (r_{\rm eff}) \, c^2 \, dt_{\rm eff}^2 + \frac{D (r_{\rm eff})}{A 
(r_{\rm eff})} \ d r_{\rm eff}^2 + r_{\rm eff}^2 (d \theta^2 + \sin^2 \theta \, d 
\varphi^2) \, ,
\ee
is possible, if and only if the energy mapping $E_{\rm eff} = f (E_{\rm real})$ is 
given by
\be
\label{eq2.16}
\frac{E_{\rm eff}}{\mu \, c^2} = \frac{E_{\rm real}^2 - m_1^2 \, c^4 - m_2^2 \, c^4}{2 
\, m_1 \, m_2 \, c^4} \, .
\ee

Remarkably, the simple energy map (\ref{eq2.16}) (which is here determined by our 
requirements) coincides with the energy map introduced in several other investigations 
\cite{BIZ}, \cite{DIS98} (and is simply related to the one defined a priori in 
\cite{T1,T2,FT01}).

Recently, the problem of mapping the extremely complicated real two-body 3~PN 
Hamiltonian (\ref{eq2.6}) onto an effective one-body dynamics has been solved 
\cite{DJS2}. Again the result is remarkably simple, though less simple than at the 
2~PN level. Indeed, it was found\footnote{In fact, \cite{DJS2} found that it was 
possible to map the real dynamics onto the geodesic dynamics of a test particle. 
However, both the effective metric and the modified energy map needed for this 
representation are rather complicated. It was felt that it is more convincing to keep 
a simple effective metric, and a simple energy map, but to relax the constraint of 
{\em geodesic} motion.} that the effective one-body dynamics was given by an 
Hamilton-Jacobi equation of the form (with $p_{\alpha}^{\rm eff} = \partial S / 
\partial \, x_{\rm eff}^{\alpha}$)
\be
\label{eq2.17}
0 = \mu^2 + g_{\rm eff}^{\alpha \beta} \, p_{\alpha}^{\rm eff} \, p_{\beta}^{\rm eff} 
+ Q_4 (p^{\rm eff}) \, ,
\ee
with a simple (spherically symmetric) effective metric of the form (\ref{eq2.15}) and 
some additional quartic-in-momenta contribution $Q_4 (p)$. Remarkably, it was found 
that, at the 3~PN level, the energy mapping is again uniquely determined to be the 
simple relation (\ref{eq2.16}). As for the metric coefficients of the (covariant) 
effective metric $g_{\alpha \beta}^{\rm eff}$, and the quartic terms $Q_4 (p)$, they 
were found to be
\be
\label{eq2.18}
A (r) = 1 - 2 \, \widehat u + 2 \, \nu \, {\widehat u}^3 + a_4 (\nu) \, {\widehat u}^4 
\, ,
\ee
\be
\label{eq2.19}
D (r) = 1 - 6 \, \nu \, {\widehat u}^2 + 2 (3 \nu - 26) \, \nu \, {\widehat u}^3 \, ,
\ee
\be
\label{eq2.20}
\frac{Q_4 (p)}{\mu^2} = 2 (4 - 3 \nu) \, \nu \, {\widehat u}^2 (\mbox{\boldmath$n$} 
\cdot \widehat{\mbox{\boldmath$p$}})^4 \, ,
\ee
where $\widehat u \equiv GM / r$, $\widehat{\mbox{\boldmath$p$}} = 
\mbox{\boldmath$p$} / \mu$ and
\be
\label{eq2.21}
a_4 (\nu) = \left( \frac{94}{3} - \frac{41}{32} \, \pi^2 + 2 \, \omega_s \right) \nu 
\, .
\ee
As we said above, the correct value of $\omega_s$ has been recently found \cite{DJSd}
to be simply $\omega_s = 0$. We shall find, however, convenient to keep $\omega_s$ as
a free parameter in order to assess the quantitative importance of 3PN effects.

Let us emphasize again the streamlined nature of the effective one-body description of 
the orbital dynamics. Successively, as the PN order increases, one can say that: (i) 
the 6 terms of the Newtonian plus first post-Newtonian relative Hamiltonian 
(\ref{eq2.3}), (\ref{eq2.4}) can be mapped (via (\ref{eq2.16}) and a canonical 
transformation of $(\mbox{\boldmath$x$} , \mbox{\boldmath$p$})$) onto geodesic motion 
in a Schwarzschild spacetime of mass $M$ (i.e. $A_{1 \, {\rm PN}} = 1 - 2 \, {\widehat 
u}$, $D_{1 \, {\rm PN}} = 1$), (ii) to take into account the 7 additional terms 
entering $H_{2 \, {\rm PN}}$, Eq.~(\ref{eq2.5}), it is enough to add $+ \, 2 \, \nu \, 
{\widehat u}^3$ to $A (r)$ and $- \, 6 \, \nu \, {\widehat u}^2$ to $D (r)$, and (iii) to 
take into account the 11 additional terms entering $H_{3 \, {\rm PN}}$, it is enough 
to further add $+ \, a_4 (\nu) \, {\widehat u}^4$ to $A (r)$ and $+ \, 2 (3 \nu - 26) \, 
\nu 
\, {\widehat u}^3$ to $D(r)$, and to add the simple quartic term (\ref{eq2.20}) to the 
mass-shell condition (\ref{eq2.17}). Note that the effective one-body dynamics is a 
``deformation'' of a geodesic dynamics for a particle of mass $\mu$ in a Schwarzschild 
spacetime of mass $M$, with deformation parameter the symmetric mass ratio $\nu = \mu 
/ M = m_1 \, m_2 / (m_1 + m_2)^2$ ($0 < \nu \leq \frac{1}{4}$, with $\nu$ being small 
if $m_1 \ll m_2$ or $m_2 \ll m_1$, and reaching its maximal value of $\frac{1}{4}$ 
when $m_1 = m_2$). Note also that, at this stage, we have not yet introduced any 
particular resummation technique. The effective quantities 
(\ref{eq2.18})--(\ref{eq2.20}) are still given as straightforward $PN$ expansions in 
powers of $\widehat u = GM / c^2 \, r$. However, this already means an 
appreciable gain over the original $PN$ expansions, Eqs.~(\ref{eq2.3})--(\ref{eq2.6}). 
Indeed, there are far less terms in the ``effective'' PN expansions and they 
generically have significantly smaller coefficients (which are now all multiplied by 
$\nu < 1/4$). For instance, the ``radial'' potential determining the circular orbits 
is now fully encoded in the simple function $g_{00}^{\rm eff} = - A (r_{\rm eff})$ 
which differs from the test-mass (Schwarzschild) result, $A_{(\nu \, = \, 0)} = 1 - 2 
\, \widehat u$, only by two numerically smallish terms when one is above the last 
stable orbit. Indeed, when $r_{\rm eff} > 6 \, GM / c^2$, $2 \, \nu \, {\widehat 
u}^3 < 0.23\%$ and $a_4 (\nu) \, {\widehat u}^4 < 0.36\%$ (for $\omega_s = 0$). When 
working at the 2~PN level it was, in fact, found unnecessary \cite{BD99} to further 
``resum'' the effective PN expansions of $A(r)$ and $D(r)$. As they stand, they led to 
small deviations from the test-mass ($\nu \rightarrow 0$) results. However, it was 
found in \cite{DJS2} that the situation is not quite as rosy at the 3~PN level. 
Because of the largish coefficient in Eq.~(\ref{eq2.21}),
$\frac{94}{3} - \frac{41}{32} \, \pi^2 \simeq 18.688$, the additional term $+ \, a_4 (\nu) 
\, \widehat{u}^4$ in Eq.~(\ref{eq2.18}) significantly modifies the qualitative 
behaviour of the metric coefficient $A(r)$ for $r_{\rm eff} < 6~GM$. In 
particular, when $\omega_s = 0$,
 the straightforward PN-expanded function $A(r)$ no longer features 
a zero near $r_{\rm eff} = 2~GM$ for all possible values of the deformation 
parameter $0 < \nu \leq \frac{1}{4}$. As this zero (which corresponds to the 
Schwarzschild horizon) is a crucial qualitative feature of the $\nu = 0$ limit, it was 
argued in \cite{DJS2} that one should Pad\'e-resum the PN-expansion of $A(r)$ so as to 
ensure the stable presence of a similar ``horizon'' when $0 < \nu \leq \frac{1}{4}$. We 
shall also do so here, but only after having introduced the spin effects, which modify 
the radial function which is the analog of $A(r)$.

\subsection{Effective spin}\label{ssec2.2}

Let us now tackle the central task of this work: to introduce spin effects in the 
effective one-body approach. Let us first emphasize that the ambition of the
present work is somewhat limited. Our main goal is to derive a spin-dependent EOB
Hamiltonian which is physically reliable for small and moderate spins. We shall proceed
toward this goal in successive steps. First, we consider only 
 (at the lowest PN order where they enter) the interactions which are {\em 
linear} in the spins. Then, we shall incorporate (to some approximation) the interactions 
which are quadratic in the spins.
Contrary to the case of the spin-independent interactions where many years of work have
yielded high-PN-accuracy information which allowed one to refine and test the EOB 
approach, one does not have in hand enough information for gauging the reliability
of the spin-dependent interactions in the case of fast spins. As a consequence, we
will be able to trust the presently introduced EOB Hamiltonian only when spin 
effects are not too large.  As we said in the Introduction, one will
need new information (either from numerical relativity, or from improved analytical
methods) to find a reliable  form of the EOB Hamiltonian for large spins.

We have written down in Eq.~(\ref{eq2.9}) above the contribution to the real, 
PN-expanded, two-body Hamiltonian which is linear in the two spins. Our first proposal 
is to map this contribution to the spin-orbit coupling of a (spinless) effective 
particle moving in a suitably ``spinning'' effective metric, i.e. some type of 
generalized Kerr metric. If we formally consider $\mbox{\boldmath$S$}_1$ and 
$\mbox{\boldmath$S$}_2$ as deformation parameters (on top of the basic {\em orbital} 
deformation parameter $\nu$), the effective dynamics we are looking for should be a 
``spin deformation'' of the currently most accurate orbital dynamics, as described by 
the 3~PN effective dynamics Eqs.~(\ref{eq2.15})--(\ref{eq2.21}). In particular, we 
should keep the non-geodesic Hamilton-Jacobi equation (\ref{eq2.17}). Note that the 
spin effects (notably Eq.~(\ref{eq2.9})) break the $O(3)$ symmetry of the orbital 
interaction. At the quantum level, this means that spin interactions lift the 
degeneracy ($=$ lack of dependence on the ``magnetic'' quantum number $L_z / \hbar$) of 
the 
orbital energy states. This shows that the effective (co)metric $g_{\rm eff}^{\alpha 
\beta}$ entering 
Eq.~(\ref{eq2.17}) should no longer be spherically symmetric, but should contain 
special directions linked to the spins. For generality, let us first consider an 
arbitrary (time-independent) effective cometric
\be
\label{eq2.22}
g_{\rm eff}^{\alpha\beta} \, p_{\alpha} \, p_{\beta} =  g_{\rm eff}^{00} \, p_0^2 + 2 
\, g_{\rm eff}^{0i} \, p_0 \, p_i + g_{\rm eff}^{ij} \, p_i \, p_j \, .
\ee
Let us define
\be
\label{eq2.23}
\alpha \equiv (-g_{\rm eff}^{00})^{-1/2} \, , \ \beta^i \equiv \frac{g_{\rm 
eff}^{0i}}{g_{\rm eff}^{00}} \, , \ \gamma^{ij} \equiv g_{\rm eff}^{ij} - \frac{g_{\rm 
eff}^{0i} \, g_{\rm eff}^{0j}}{g_{\rm eff}^{00}} \, ,
\ee
i.e.
\be
\label{eq2.24}
g_{\rm eff}^{00} = - \frac{1}{\alpha^2} \, , \ g_{\rm eff}^{0i} = - 
\frac{\beta^i}{\alpha^2} \, , \ g_{\rm eff}^{ij} = \gamma^{ij} - \frac{\beta^i \, 
\beta^j}{\alpha^2} \, .
\ee

The effective energy $E_{\rm eff} \equiv -p_0^{\rm eff}$ is conserved (because of the 
assumed stationarity of $g_{\rm eff}^{\alpha\beta}$). Using the parametrization 
(\ref{eq2.23}), Eq.~(\ref{eq2.17}) reads
\be
\label{eq2.25}
(E_{\rm eff} - \beta^i \, p_i)^2 = \alpha^2 \, [\mu^2 + \gamma^{ij} \, p_i \, p_j + Q_4 
(p)] \, .
\ee

Solving Eq.~(\ref{eq2.25}) for $E_{\rm eff}$ (using the fact that $Q_4 (p)$, 
Eq.~(\ref{eq2.20}), depends only on the $p_i$'s) yields the effective Hamiltonian
\be
\label{eq2.26}
E_{\rm eff} = H_{\rm eff} (\mbox{\boldmath$x$} , \mbox{\boldmath$p$} , \ldots) = 
\beta^i \, p_i + \alpha \, \sqrt{\mu^2 + \gamma^{ij} \, p_i \, p_j + Q_4 (p_i)} \, ,
\ee
where we have suppressed, for readability, the labels ``eff'' on the orbital phase 
space variables $\mbox{\boldmath$x$}^{\rm eff}$, $\mbox{\boldmath$p$}^{\rm eff}$. The 
ellipsis in the arguments of $H_{\rm eff}$ are added to remind us that $H_{\rm eff}$ 
will ultimately also depend on the spin variables $\mbox{\boldmath$S$}_1$, 
$\mbox{\boldmath$S$}_2$, which enter the metric coefficients $\alpha$, $\beta^i$, 
$\gamma^{ij}$ as parameters.

We also assume that Eq.~(\ref{eq2.16}) (which was found to hold at 1~PN, 2~PN and 3~PN) 
still holds. Solving it for the real energy $E_{\rm real}$ in terms of the effective 
one finally yields the real Hamiltonian
\be
\label{eq2.27}
E_{\rm real} = H_{\rm real} (\mbox{\boldmath$x$} , \mbox{\boldmath$p$} , \ldots) = M \, 
\sqrt{1 + 2 \, \nu \left( \frac{H_{\rm eff} - \mu}{\mu} \right)} \, .
\ee

We recall that, at the linearized level and at the lowest PN order, the addition of a 
spin $\mbox{\boldmath$S$}_{\rm eff}$ onto an initially spherical symmetric metric leads 
to an off-diagonal term in the metric:
\be
\label{eq2.28}
\beta^i \simeq - g^{0i} \simeq - \, g_{0i} \simeq + \frac{2G}{r^3} \ \varepsilon^{ijk} \, 
S_{\rm 
eff}^j \, x^k \, .
\ee

Inserting this term in (\ref{eq2.26}), and expanding (\ref{eq2.27}) in a PN series 
yields, as leading spin-orbit coupling (linearized in $S_{\rm eff}$ and taken to formal 
order ${\cal O} (1/c^2)$) in $H_{\rm real}$, the term
\be
\label{eq2.29}
\delta_{S_{\rm eff}} \, H_{\rm real} \simeq \beta^i \, p_i \simeq \frac{2G}{c^2 \, r^3} 
\ \varepsilon^{ijk} \, p_i \, S_{\rm eff}^j \, x^k \, .
\ee
This term can exactly reproduce the leading\footnote{We use here the fact that the real 
phase-space coordinates $\mbox{\boldmath$x$}^{\rm real}$, $\mbox{\boldmath$p$}^{\rm 
real}$ differ only by ${\cal O} (1/c^2)$ from the effective ones entering $H_{\rm eff}$ 
\cite{BD99}.} two-body spin-orbit coupling (\ref{eq2.9}) if we define
\be
\label{eq2.30}
\mbox{\boldmath$S$}^{\rm eff} \equiv \sigma_1 \, \mbox{\boldmath$S$}_1 + \sigma_2 \, 
\mbox{\boldmath$S$}_2 \, ,
\ee
with
\be
\label{eq2.31}
\sigma_1 \equiv 1 + \frac{3}{4} \, \frac{m_2}{m_1} \ , \quad \sigma_2 \equiv 1 + 
\frac{3}{4} \, \frac{m_1}{m_2} \, .
\ee

\subsection{A deformed Kerr metric}\label{ssec2.3}

Remembering that the main message of the effective one-body method is that the orbital 
dynamics of two comparable-mass black holes can be described in terms of a slightly 
deformed (with deformation parameter $\nu$) Schwarzschild metric, we expect that the 
orbital-plus-spin dynamics of two black holes can be described in terms of some 
deformation of the Kerr metric. In other words, we are expecting that not only the 
effects linear in the spins, such as Eq.~(\ref{eq2.9}), but also the spin-dependent 
non-linear effects, can be described in terms of some deformed, effective Kerr metric. 
At this stage it is therefore very natural to construct a suitable ``deformed Kerr 
metric'' which combines the orbital deformations (\ref{eq2.18}), (\ref{eq2.19}) with 
the full spin effects linked to the ``effective spin'' (\ref{eq2.30}). After 
constructing this deformed Kerr metric, we shall a posteriori check that it 
approximately incorporates the expected two-body interactions which are {\em quadratic} 
in the spins.

Let us start from the simplest form of the Kerr cometric, underlying its separability 
properties \cite{carter}
\begin{eqnarray}
\label{eq2.32}
g_{\rm Kerr}^{\alpha\beta} \, p_{\alpha} \, p_{\beta} && \, = \frac{1}{r^2 + a^2 \cos^2 
\theta} \nonumber \\
\times&&\left[ \Delta_K (r) \, p_r^2 + p_{\theta}^2 + \frac{1}{\sin^2 \theta} \, 
(p_{\varphi} 
+ a \sin^2 \theta \, p_t)^2 - \frac{1}{\Delta_K (r)} \, ((r^2 + a^2) \, p_t + a \, 
p_{\varphi})^2 \right] \, ,
\end{eqnarray}
with $\Delta_K (r) = r^2 - 2Mr + a^2$. In the non-spinning limit $(a \rightarrow 0)$ 
the coefficients of $p_r^2$ and $p_t^2$ become, respectively, $\Delta_K (r) / r^2$ and $- 
r^2 / \Delta_K (r)$. However, we know that in this limit we should get (from 
(\ref{eq2.15})) $A(r) / D(r)$ and $- 1 / A(r)$, respectively. It is therefore very 
natural to generalize the Kerr metric (\ref{eq2.32}) (while still keeping its 
separability properties) by assuming that the coefficients of the first and last terms 
in the square brackets of (\ref{eq2.32}) involve two different functions of $r$, say 
$\Delta_r (r)$ and $-1 / \Delta_t (r)$, whose product reduces to $-1/D(r)$ when $a 
\rightarrow 0$. This reasoning leads us, as simplest\footnote{We leave untouched the 
dependence on $a$ to ensure that, when $GM \rightarrow 0$ with $a$ being fixed, 
the metric $g_{\rm eff}^{\alpha\beta}$ be Minkowski in disguise.} possibility for 
combining spin effects with orbital effects, to postulating that the effective metric 
entering (\ref{eq2.17}) has the form
\begin{eqnarray}
\label{eq2.33}
g_{\rm eff}^{\alpha\beta}  \, p_{\alpha} \, p_{\beta} && \, = \frac{1}{r^2 + a^2 \cos^2 
\theta} \nonumber \\
\times&&\left[ \Delta_r (r) \, p_r^2 + p_{\theta}^2 + \frac{1}{\sin^2 \theta} \, 
(p_{\varphi} 
+ a \sin^2 \theta \, p_t)^2 - \frac{1}{\Delta_t (r)} \, ((r^2 + a^2) \, p_t + a \, 
p_{\varphi})^2 \right] \, ,
\end{eqnarray}
with
\be
\label{eq2.34}
\Delta_r^{\rm PN} (r) = \frac{r^2 \, A^{\rm PN} (r) + a^2}{D^{\rm PN} (r)} \ , \quad 
\Delta_t^{\rm PN} (r) = r^2 \, A^{\rm PN} (r) + a^2 \, .
\ee
Here, the superscripts ``PN'' indicate that, at this stage, we are only comparing PN 
expansions. We already know from the 3~PN study of \cite{DJS2} that this is 
unsatisfactory because it tends to change the qualitative behaviour of the radial 
functions, and, in particular, the presence of a horizon in the metric (\ref{eq2.33}). 
To get a regular horizon in Eq.~(\ref{eq2.33}) we need the two functions $\Delta_t (r)$ 
and $\Delta_r(r)$ to have a zero at the same value of $r$. The simplest (and most 
robust) way of ensuring this is (as discussed in \cite{DJS2}) to define them as
\be
\label{eq2.35}
\Delta_t (r) \equiv r^2 \, P_3^1 \left[ A^{\rm PN} (r) + \frac{a^2}{r^2} \right] \ , 
\quad \Delta_r (r) \equiv \Delta_t (r) \left( \frac{1}{D(r)} \right)^{\rm PN} \, .
\ee
Here, $P_m^n [f^{\rm PN} (u)]$, with $u \equiv 1/r$, denotes the $N_n (u) / D_m (u)$ 
Pad\'e of a certain PN-expanded function $f^{\rm PN} (u) = c_0 + c_1 \, u + c_2 \, u^2 
+ \cdots + c_{n+m} \, u^{n+m}$ ($N_n (u)$ and $D_m (u)$ being polynomials in $u$ of 
degrees $n$ and $m$, respectively).
We do not write down the (uniquely defined) explicit expression of
\begin{eqnarray}
\bar A (u) \equiv P_3^1 \, [A^{\rm PN} (u) + a^2 \, u^2] &= &P_3^1 \, [1 - 2 \widehat{u} + 
\widehat{a}^2 \, \widehat{u}^2 + 2 \nu \, \widehat{u}^3 + a_4 (\nu) \, \widehat{u}^4] 
\nonumber \\
&= &\frac{1 + n_1 \, \widehat{u}}{1 + d_1 \, \widehat{u} + d_2 \, \widehat{u}^2 + d_3 \, 
\widehat{u}^3} \nonumber
\end{eqnarray}
(where $\widehat{u} = GM / r$, $\widehat{a} = a / GM$) because: (i) it is rather 
complicated and not illuminating, and (ii) modern algebraic manipulators compute it 
directly from its Pad\'e definition.

In the definition of $\Delta_r (r)$ (which is less 
important than that of $\Delta_t (r)$) we have factorized the Pad\'eed $\Delta_t (r)$ 
and assumed that it was enough to work with the non-resummed PN-expansion of the 
inverse of the $D$-function, i.e. (from (\ref{eq2.19}))
\be
\label{eq2.36}
(D^{-1} (r))^{\rm PN} \equiv 1 + 6 \, \nu \, \widehat{u}^2 + 2 \, (26 - 3 \nu) \, \nu 
\, \widehat{u}^3 \, .
\ee
If the need arises, it would be easy to define improved (resummed) versions of $D^{-1} 
(r)$. Because of the positive coefficients in (\ref{eq2.36}) the present definition 
does not interfere (as would the consideration of $(D^{\rm PN} (r))^{-1}$) with the 
desired feature of having a simple zero in $\Delta_r (r)$ located at the same value as 
that in $\Delta_t (r)$.

Finally, we shall see later that there are some advantages in defining the 
quartic-in-momenta contribution $Q_4 (p)$ in the following (deformed) way
\be
\label{eq2.36new}
Q_4 (p) = \frac{1}{\mu^2} \, 2 \, (4 - 3 \nu) \, \nu \ \frac{(GM)^2}{r^2 + a^2 \cos^2 
\theta} \ (\mbox{\boldmath$n$} \cdot \mbox{\boldmath$p$})^4 \, .
\ee

Eq.~(\ref{eq2.33}) defines $g_{\rm eff}^{\alpha\beta}$ only with respect to some 
(instantaneous) polar coordinate system with $z$-axis aligned with the effective spin 
(\ref{eq2.30}). Such a coordinate system cannot be used for describing the evolution of 
two gravitationally interacting spinning black holes. Indeed, we expect (and shall 
check below) that the total real Hamiltonian imposes some type of precession motion for 
$\mbox{\boldmath$S$}_1$, $\mbox{\boldmath$S$}_2$ and therefore 
$\mbox{\boldmath$S$}_{\rm eff}$. To get the full dynamics of the system we need to 
rewrite $g_{\rm eff}^{\alpha\beta}$ in a general, cartesian-like coordinate system. 
This is achieved by explicitly introducing, besides $n^i \equiv x^i / r$, the 
quantities $(S_{\rm eff} \equiv (\delta_{ab} \, S_{\rm eff}^a \, S_{\rm eff}^b)^{1/2})$
\be
\label{eq2.37}
s^i \equiv \frac{S_{\rm eff}^i}{S_{\rm eff}} \ , \quad a \equiv \frac{S_{\rm eff}}{M} \ 
, \quad \cos \theta \equiv n^i \, s^i \ , \quad \rho^2 \equiv r^2 + a^2 \cos^2 \theta 
\, . 
\ee
This leads to the following, cartesian-like, effective metric
\be
\label{eq2.38}
- \rho^2 \, g_{\rm eff}^{00} = \frac{(r^2 + a^2)^2 - a^2 \, \Delta_t \sin^2 
\theta}{\Delta_t} \, ,
\ee
\be
\label{eq2.39}
\rho^2 \, g_{\rm eff}^{0i} = - \frac{a \, [r^2 + a^2 - \Delta_t]}{\Delta_t} \ 
(\mbox{\boldmath$s$} \times \mbox{\boldmath$x$})^i \, ,
\ee
\be
\label{eq2.40}
\rho^2 \, g_{\rm eff}^{ij} = \Delta_r \, n^i \, n^j + r^2 \, (\delta^{ij} - n^i \, n^j) 
- \frac{a^2}{\Delta_t} \ (\mbox{\boldmath$s$} \times \mbox{\boldmath$x$})^i \, 
(\mbox{\boldmath$s$} \times \mbox{\boldmath$x$})^j \, ,
\ee
from which follows
\be
\label{eq2.41}
\alpha = \sqrt{\frac{\rho^2 \, \Delta_t}{(r^2 + a^2)^2 - a^2 \, \Delta_t \sin^2 
\theta}} \, ,
\ee
\be
\label{eq2.42}
\beta^i = \frac{a \, [r^2 + a^2 - \Delta_t] \, (\mbox{\boldmath$s$} \times 
\mbox{\boldmath$x$})^i}{(r^2 + a^2)^2 - a^2 \, \Delta_t \sin^2 \theta} \, ,
\ee
\be
\label{eq2.43}
\gamma^{ij} = g_{\rm eff}^{ij} + \frac{\beta^i \, \beta^j}{\alpha^2} \, .
\ee
Note that near the ``horizon'', i.e. as $\Delta_t \rightarrow 0$, the quantity $\alpha$ 
tends to zero like $\sqrt{\Delta_t}$, while $\beta^i$ and $\gamma^{ij}$ have finite 
limits. [The singular last term on the right-hand side of Eq.~(\ref{eq2.40}) is 
cancelled near the horizon by the contribution $+ \beta^i \, \beta^j / \alpha^2$ to 
$\gamma^{ij}$, 
Eq.~(\ref{eq2.43}).]

Finally, the spin-dependent, real two-body Hamiltonian $H_{\rm real} 
(\mbox{\boldmath$x$} , \mbox{\boldmath$p$} , \mbox{\boldmath$S$}_1 , 
\mbox{\boldmath$S$}_2)$ is defined by ($\widehat{p}_i \equiv p_i / \mu$, $\widehat 
u_{\rho} 
\equiv GM / \sqrt{\rho^2}$, $n^i \equiv x^i / r$)
\begin{eqnarray}
\label{eq2.44}
H_{\rm real} (\mbox{\boldmath$x$} , \mbox{\boldmath$p$} , \mbox{\boldmath$S$}_1 , 
\mbox{\boldmath$S$}_2) \nonumber \\
\equiv &M \sqrt{1 + 2\nu \left( \beta^i \, \widehat{p}_i + \alpha \sqrt{1 + \gamma^{ij} 
\, \widehat{p}_i \, \widehat{p}_j + 2 \, (4-3\nu) \, \nu \,  \widehat{u}_{\rho}^2 (n^i \, 
\widehat{p}_i)^4} - 1 \right)} \, ,
\end{eqnarray}
where we recall that the basic effective Kerr spin vector is defined by
\be
\label{eq2.45}
M \, a \, s^i \equiv S_{\rm eff}^i \equiv \sigma_1 \, S_1^i + \sigma_2 \, S_2^i \, ,
\ee
with $\sigma_1$ and $\sigma_2$ defined in Eq.~(\ref{eq2.31}). The phase-space 
coordinates appearing in this Hamiltonian are the effective ones $(x_{\rm eff}^i , 
p_i^{\rm eff})$. They differ \cite{BD99,DJS2} by ${\cal O} (1/c^2)$ terms from the 
coordinates used in usual PN calculations, such as ADM ones. The evolution equations 
defined by the Hamiltonian (\ref{eq2.44}) are obtained by the Poisson bracket equations 
(\ref{eq2.10})--(\ref{eq2.13}). Before discussing them let us show how the Hamiltonian 
(\ref{eq2.44}) contains spin-quadratic effects of the good sign and magnitude.

\subsection{Effects quadratic in the spins}\label{ssec2.4}

Note first that if we introduce the ``non relativistic'' effective Hamiltonian $H_{\rm 
eff}^{\rm NR} \equiv H_{\rm eff} - \mu \, c^2$, and similarly $H_{\rm real}^{\rm NR} 
\equiv H_{\rm real} - M \, c^2$, one has
\be
\label{eq2.46}
H_{\rm eff}^{\rm NR} = H_{\rm real}^{\rm NR} \left( 1 + \frac{1}{2} \ \frac{H_{\rm 
real}^{\rm NR}}{M \, c^2} \right) \, .
\ee
Therefore, if we are interested in the leading PN approximation to any additional term 
in $H_{\rm real}^{\rm PN}$, one can neglect the (${\cal O} (1/c^2)$ smaller) difference 
between $H_{\rm eff}$ and $H_{\rm real}$. By this argument, the leading PN 
approximation to the term linear in the spins is
\be
\label{eq2.47}
H_{{\rm real} \, S} \simeq H_{{\rm eff} \, S} \simeq \beta^i \, p_i = \frac{a \, [r^2 + 
a^2 - \Delta_t]}{(r^2 + a^2)^2 - a^2 \Delta_t \sin^2 \theta} \ (\mbox{\boldmath$s$} 
\times \mbox{\boldmath$x$})^i \, p_i \, .
\ee
We write it explicitly in the form in which it appears in our Hamiltonian for the 
reader to see how the term (\ref{eq2.29}) is generated. [The important feature here 
being that $r^2 + a^2 - \Delta_t \simeq 2 \, GM \, r$ at the leading PN 
approximation.]

Let us now consider the interaction terms in $H_{\rm real}$ or $H_{\rm eff}$ which are 
quadratic in the spins, and therefore quadratic in the Kerr-like parameter $a$, 
Eq.~(\ref{eq2.45}). First, one should remember that most terms of order $a^2$, as they 
appear in the effective metric (\ref{eq2.38})--(\ref{eq2.40}), do not directly 
correspond to physical effects proportional to $\mbox{\boldmath$S$}_{\rm eff}^2$. 
Indeed, we are using here Boyer-Lindquist-type coordinates which differ, even in the 
flat space limit $GM \rightarrow 0$, from usual (flat-space, cartesian-like) 
coordinates by terms of order ${\cal O} (a^2)$. As we are interested in the leading PN 
effects quadratic in the spins, we can view the Kerr-like metric 
(\ref{eq2.38})--(\ref{eq2.40}) as a deformation, by the $a$-dependent terms, of the 
Schwarzschild metric (which is the leading PN version of the orbital effective metric). 
We then expect that the leading physical effects quadratic $a$ will be those linked to 
the $a$-dependent quadrupole moment deformation of the Schwarzschild metric. The 
quadrupole moment of the Kerr metric (which coincides with our metric when we neglect 
additional 2~PN fractional corrections) has been determined to be \cite{thorne}
\be
\label{eq2.48}
Q_{ij} = -M \, a^2 \, s_i \, s_j + \frac{1}{3} \ M a^2 \, \delta_{ij} \, .
\ee
This corresponds to an additional term in the interaction Hamiltonian equal to
\be
\label{eq2.49}
H_Q^{\rm real} \simeq H_Q^{\rm eff} \simeq - \frac{1}{2} \, \mu \, Q_{ij} \, 
\partial_{ij} \, \frac{1}{r} = + \frac{1}{2} \, \mu \, Ma^2 \, s^i \, s^j \, 
\partial_{ij} \, \frac{1}{r} \, .
\ee
In terms of the effective spin this reads (in standard units)
\be
\label{eq2.50}
H_Q^{\rm real} \simeq + \frac{1}{2} \, \frac{G}{c^2} \, \frac{\mu}{M} \ S_{\rm eff}^i 
\, 
S_{\rm eff}^j \, \partial_{ij} \, \frac{1}{r} \, .
\ee
Such is the prediction from our Hamiltonian. Let us now compare it to the expected real 
two-body, spin-quadratic effects. As sketched in Eq.~(\ref{eq2.8}) there are several 
sources of spin-quadratic effects. At leading PN order, it is enough to consider: (i) 
the term $\propto \, m_2 \, S_1^2 / m_1$ which arises because of the interaction of the 
monopole $m_2$ with the spin-induced quadrupole moment of the spinning black hole of 
mass $m_1$, (ii) the term $\propto \, m_1 \, S_2^2 / m_2$ obtained by exchanging $1 
\rightarrow 2$, and (iii) the term $\propto \, S_1 \, S_2$ coming from the direct, 
one-graviton interaction between the two spin-dipoles. The first term is obtained by 
relabelling the result (\ref{eq2.50}) by $\mu \rightarrow m_2$, $M \rightarrow m_1$, 
$S_{\rm eff}^i \rightarrow S_1^i$. Therefore the sum of (i) and (ii) reads
\be
\label{eq2.51}
H_{S_1 S_1} + H_{S_2 S_2} \simeq + \frac{1}{2} \, \frac{G}{c^2} \left( \frac{m_2}{m_1} 
\, S_1^i \, S_1^j + \frac{m_1}{m_2} \, S_2^i \, S_2^j \right) \, \partial_{ij} \, 
\frac{1}{r} \, .
\ee
The term (iii) has been computed in \cite{BGH66,BOC70} and reads
\be
\label{eq2.52}
H_{S_1 S_2} \simeq + \frac{G}{c^2 \, r^3} \, S_1^i \, S_2^j \, (3 \, n^i \, n^j - 
\delta^{ij}) = + \frac{G}{c^2} \, S_1^i \, S_2^j \, \partial_{ij} \, \frac{1}{r} \, .
\ee
It is easily checked that the sum of (\ref{eq2.51}) and (\ref{eq2.52}), say $H_{SS} 
\equiv H_{S_1 S_1} + H_{S_2 S_2} + H_{S_1 S_2}$, can be written as
\be
\label{eq2.53}
H_{SS} \simeq \frac{1}{2} \, \frac{G}{c^2} \, \frac{\mu}{M} \, S_0^i \, S_0^j \, 
\partial_{ij} \, \frac{1}{r} \, ,
\ee
with $S_0^i/M \equiv a_0^i \equiv a_1^i + a_2^i \equiv S_1^i/m_1 + S_2^i/m_2$,
i.e. explicitly,
\be
\label{eq2.54}
S_0^i \equiv \left(1 + \frac{m_2}{m_1} \right) \, S_1^i + \left(1 + \frac{m_1}{m_2} 
\right) \, S_2^i \, .
\ee
The result (\ref{eq2.53}), (\ref{eq2.54}) is remarkably similar to the prediction 
(\ref{eq2.50}) (with (\ref{eq2.30}), (\ref{eq2.31})). The only discrepancy is a 25\% 
difference in the coefficient of the mass ratios in the definition of the effective 
spin. Though there might be physical situations where this smallish difference might play 
a significant role, we think that in most cases where one will be entitled to trust the 
approximate spin-dependent EOB Hamiltonian introduced here
 this difference will not matter. Indeed, because of the 
partially ad hoc way in which we constructed our deformed Kerr metric, we cannot trust 
our predictions beyond the domain where spin effects are moderate corrections to 
orbital effects. However, it is useful to incorporate in a qualitatively correct manner 
the non-linear spin effects. This is what our prescription achieves. For instance: (i) 
in the limit where, say, $m_2 \ll m_1$ (and $\vert \mbox{\boldmath$S$}_2 \vert \leq 
m_2^2$) (\ref{eq2.50}) and (\ref{eq2.53}) become equivalent, or (ii) in the case where 
$\mbox{\boldmath$S$}_1$ and $\mbox{\boldmath$S$}_2$ are parallel (in the same 
direction), (\ref{eq2.50}) and (\ref{eq2.53}) differ only by a numerical factor which 
is near one for all mass ratios. 

It might, however, be useful to define another Hamiltonian, say $H'_{\rm real}$, which: 
(a) 
reduces (like $H_{\rm real}$) to the Kerr one in the test-mass (and test-spin) limit, (b) 
contains (like $H_{\rm real}$) the 
spin-orbit 
terms (\ref{eq2.9}), and (c) contains the exact spin-spin terms (\ref{eq2.53}) (instead of 
their ``25\%'' approximation (\ref{eq2.50}) contained in $H_{\rm real}$). A simple way to 
do that is to define $H'_{\rm real} \equiv M \, \sqrt{1 + 2 \, \nu (H'_{\rm eff} - \mu) / 
\mu}$ with a modified effective Hamiltonian defined as the sum of (\ref{eq2.26}), written 
with the replacement $\mbox{\boldmath$S$}_{\rm eff} \rightarrow \mbox{\boldmath$S$}_0$, 
and 
of an additional spin-orbit interaction term, $\Delta \beta^i \, p^i$, with $\Delta 
\beta^i$ proportional to the difference $\sigma^i \equiv S_{\rm eff}^i - S_0^i$:
\be
\label{eq2.56}
H'_{\rm eff} (\mbox{\boldmath$x$} , \mbox{\boldmath$p$} , \mbox{\boldmath$S$}_1 , 
\mbox{\boldmath$S$}_2) = H_{\rm eff} (\mbox{\boldmath$x$} , \mbox{\boldmath$p$} , 
\mbox{\boldmath$S$}_0) + \Delta H_{\rm SO} (\mbox{\boldmath$x$} , \mbox{\boldmath$p$} , 
\mbox{\boldmath$S$}_0 , \mbox{\boldmath$\sigma$}) \, ,
\ee
where (denoting $\mbox{\boldmath$a$}_0 \equiv \mbox{\boldmath$S$}_0 / M$, $\cos \theta_0 
\equiv n^i \, S_0^i / \vert \mbox{\boldmath$S$}_0 \vert$)
\be
\label{eq2.57}
\Delta H_{\rm SO} (\mbox{\boldmath$x$} , \mbox{\boldmath$p$} , \mbox{\boldmath$S$}_0 , 
\mbox{\boldmath$\sigma$}) \equiv \frac{r^2 + a_0^2 - \Delta_t (a_0)}{(r^2 + a_0^2)^2 - 
a_0^2 \, \Delta_t (a_0) \sin^2 \theta_0} \ \frac{\varepsilon^{ijk} \, p_i \, \sigma^j \, 
x^k}{M} \, 
,
\ee
with
\be
\label{eq2.58}
\sigma^i \equiv S_{\rm eff}^i - S_0^i \equiv - \frac{1}{4} \left( \frac{m_2}{m_1} \ S_1^i 
+ 
\frac{m_1}{m_2} \ S_2^i  \right) \, .
\ee

The consideration of the new Hamiltonian $H'_{\rm real}$ would considerably complicate 
(even at the qualitative level) the discussion of the following section. As we are not 
sure 
that this complication really entails a better {\em quantitative} description of spin 
effects, when these become important, we shall, in the following, content ourselves with 
studying the consequences of the simpler (though slightly less ``accurate'') Hamiltonian 
$H_{\rm real}$, Eq.~(\ref{eq2.44}) with (\ref{eq2.45}). However, we mention that it might 
be useful to consider simultaneously $H_{\rm real}$ and $H'_{\rm real}$, and to trust 
their 
predictions only in the cases where they differ only by a slight amount. This gives a 
useful measure of the domain of validity of the present 
spin-dependent effective-one-body approach.


\section{Dynamics of two spinning black holes}\label{sec3}

\subsection{Equations of motion and exact or approximate first 
integrals}\label{ssec3.1}

In the previous section we have explicitly constructed an Hamiltonian $H_{\rm real} 
(\mbox{\boldmath$x$} , \mbox{\boldmath$p$} , \mbox{\boldmath$S$}_1 , 
\mbox{\boldmath$S$}_2)$ describing (to some approximation) the (conservative part of 
the) gravitational interaction of two spinning black holes in the center-of-mass frame 
of the binary system. In the present section, we shall describe some consequences of 
this Hamiltonian. Let us start by writing down explicitly the evolution equations for 
all the dynamical variables. From the basic PB's (\ref{eq2.10})--(\ref{eq2.14}) we get
\be
\label{eq3.1}
\frac{dx^i}{dt} = \{ x^i , H_{\rm real} \} = + \frac{\partial \, H_{\rm real}}{\partial 
\, p_i} \, ,
\ee
\be
\label{eq3.2}
\frac{dp_i}{dt} = \{ p_i , H_{\rm real} \} = - \frac{\partial \, H_{\rm real}}{\partial 
\, x^i} \, ,
\ee
\be
\label{eq3.3}
\frac{dS_1^i}{dt} = \{ S_1^i , H_{\rm real} \} = \varepsilon^{ijk} \,  \frac{\partial 
\, H_{\rm real}}{\partial \, S_1^j} \, S_1^k \, ,
\ee
\be
\label{eq3.4}
\frac{dS_2^i}{dt} = \{ S_2^i , H_{\rm real} \} = \varepsilon^{ijk} \,  \frac{\partial 
\, H_{\rm real}}{\partial \, S_2^j} \, S_2^k \, .
\ee
In vectorial notation, the spin evolution equations read (e.g. for the first spin)
\be
\label{eq3.5}
\frac{d \mbox{\boldmath$S$}_1}{dt} = \mbox{\boldmath$\Omega$}_1 \times 
\mbox{\boldmath$S$}_1 \, , \ \mbox{\boldmath$\Omega$}_1 \equiv \frac{\partial \, H_{\rm 
real}}{\partial \, \mbox{\boldmath$S$}_1} \, .
\ee
A first consequence of these results is that the magnitudes of the two spins are 
exactly conserved:
\be
\label{eq3.6}
\mbox{\boldmath$S$}_1^2 = {\rm const}. \, , \ \mbox{\boldmath$S$}_2^2 = {\rm const}.
\ee
Another general consequence is the exact conservation of the total angular momentum
\be
\label{eq3.7}
\mbox{\boldmath$J$} \equiv \mbox{\boldmath$L$} + \mbox{\boldmath$S$}_1 + 
\mbox{\boldmath$S$}_2 \, ,
\ee
where $L^i \equiv \varepsilon^{ijk} \, x^j \, p_k$. Indeed, it is easily checked that 
$J^i$ generates, by Poisson brackets, global rotations of all the vectorial dynamical 
quantities: $\{ J^i , V^j \} = \varepsilon^{ijk} \, V^k$ for $\mbox{\boldmath$V$} = 
\mbox{\boldmath$x$}$, $\mbox{\boldmath$p$}$, $\mbox{\boldmath$S$}_1$ or 
$\mbox{\boldmath$S$}_2$. As the Hamiltonian is a scalar constructed out of 
$\mbox{\boldmath$x$}$, $\mbox{\boldmath$p$}$, $\mbox{\boldmath$S$}_1$ and 
$\mbox{\boldmath$S$}_2$, we have
\be
\label{eq3.8}
\frac{d}{dt} \ J^i = \{ J^i , H_{\rm real} \} = 0 \, .
\ee

Therefore
\be
\label{eq3.9}
J^i = {\rm const}, \ \hbox{and, in particular,} \ \mbox{\boldmath$J$}^2 = {\rm const}.
\ee
Evidently, we have also the conservation of the total energy:
\be
\label{eq3.10}
\frac{d \, H_{\rm real}}{dt} = \{ H_{\rm real} , H_{\rm real} \} = 0 \Rightarrow H_{\rm 
real} = {\rm const}.
\ee
This closes the list of generic first integrals of the evolution. It should be noted 
that, in general, quantities such as $\mbox{\boldmath$L$}^2$ or $\mbox{\boldmath$S$}_1 
\cdot \mbox{\boldmath$S$}_2$ are not conserved in time. This means, in particular, that 
the magnitude of the effective spin, $a^2 = M^{-2} \, \mbox{\boldmath$S$}_{\rm eff}^2$, 
will not stay constant during the evolution.

Evidently, in particular situations, more quantities might be approximately conserved. 
An interesting case is that in which the spins are small enough for one to retain only 
the terms linear in them. In this approximation
\be
\label{eq3.11}
H_{\rm eff} (\mbox{\boldmath$x$}, \mbox{\boldmath$p$}, \mbox{\boldmath$S$}_1 , 
\mbox{\boldmath$S$}_2) \simeq H_0 (\mbox{\boldmath$x$}, \mbox{\boldmath$p$}) + 
\frac{[r^2 - \Delta_t^{(a \, = \, 0)}]}{M \, r^4} \ \mbox{\boldmath$L$} \cdot 
\mbox{\boldmath$S$}_{\rm eff} \, ,
\ee
where $H_0 (\mbox{\boldmath$x$}, \mbox{\boldmath$p$})$ is spherically symmetric.

Let us, more generally\footnote{For instance, we can assume Eq.~(\ref{eq3.11}) for 
$H_{\rm eff}$, but make no further approximation in computing $H_{\rm real} = f (H_{\rm 
eff})$.}, assume that $H_{\rm eff}$, as well as $H_{\rm real}$, are spherically 
symmetric, functions of $\mbox{\boldmath$x$}$ and $\mbox{\boldmath$p$}$ except for a 
dependence on the combination $\mbox{\boldmath$L$} \cdot \mbox{\boldmath$S$}_{\rm eff}$:
\be
\label{eq3.12}
H_{\rm real} = H_{\rm real} (r , p_r , \mbox{\boldmath$L$}^2 , \mbox{\boldmath$L$} \cdot 
\mbox{\boldmath$S$}_{\rm eff}) \, ,
\ee
where $p_r \equiv n^i \, p_i$ is canonically conjugated to $r$ ($\{ r , p_r \} = 1$). 
Under the assumption (\ref{eq3.12}) the angular momenta evolution equations become (with 
$({\rm LS}) \equiv \mbox{\boldmath$L$} \cdot \mbox{\boldmath$S$}_{\rm eff}$)
\be
\label{eq3.13}
\frac{d \, \mbox{\boldmath$S$}_1}{dt} = \frac{\partial \, H_{\rm real}}{\partial \, 
({\rm LS})} \ \sigma_1 \, \mbox{\boldmath$L$} \times \mbox{\boldmath$S$}_1 \, ,
\ee
\be
\label{eq3.14}
\frac{d \, \mbox{\boldmath$S$}_2}{dt} = \frac{\partial \, H_{\rm real}}{\partial \, 
({\rm LS})} \ \sigma_2 \, \mbox{\boldmath$L$} \times \mbox{\boldmath$S$}_2 \, ,
\ee
\be
\label{eq3.15}
\frac{d \, \mbox{\boldmath$L$}}{dt} = \frac{\partial \, H_{\rm real}}{\partial \, ({\rm 
LS})} \ \mbox{\boldmath$S$}_{\rm eff} \times \mbox{\boldmath$L$} \, .
\ee

These evolution equations imply not only (as in the general case) the conservation of 
$\mbox{\boldmath$J$} = \mbox{\boldmath$L$} + \mbox{\boldmath$S$}_1 + 
\mbox{\boldmath$S$}_2$, and of $\mbox{\boldmath$S$}_1^2$ and $\mbox{\boldmath$S$}_2^2$, 
but also that of:
\be
\label{eq3.16}
\mbox{\boldmath$L$}^2 = {\rm const}, \ \mbox{\boldmath$L$} \cdot 
\mbox{\boldmath$S$}_{\rm eff} = {\rm const}.
\ee
Note, however, that $\mbox{\boldmath$S$}_{\rm eff}^2$ is not conserved. Moreover, the 
radial motion is governed by the equations
\be
\label{eq3.17}
\dot r = \frac{\partial \, H_{\rm real} (r , p_r , \mbox{\boldmath$L$}^2 , 
\mbox{\boldmath$L$} \cdot \mbox{\boldmath$S$}_{\rm eff})}{\partial \, p_r} \, ,
\ee
\be
\label{eq3.18}
\dot{p}_r = - \frac{\partial \, H_{\rm real} (r , p_r , \mbox{\boldmath$L$}^2 , 
\mbox{\boldmath$L$} \cdot \mbox{\boldmath$S$}_{\rm eff})}{\partial \, r} \, .
\ee
In view of the constancy of $\mbox{\boldmath$L$}^2 = C_0$ and $\mbox{\boldmath$L$} \cdot 
\mbox{\boldmath$S$}_{\rm eff} = C_1$, we see from these equations that the function of 
$r$ and $p_r$, $H_{\rm rad} (r , p_r) = H_{\rm real} (r , p_r , C_0 , C_1)$, defines a 
reduced Hamiltonian describing the radial motion, separately from the angular degrees of 
freedom. In particular, we see (using the fact that $p_r$ enters at least quadratically
in $H_{\rm real} $)
that, under our current (approximate) assumption 
(\ref{eq3.12}), there exists a class of {\em spherical orbits}, i.e. of orbits 
satisfying 
\be
\label{eq3.19}
r = {\rm const} \, , \ p_r = 0 \, , \ \frac{\partial \, H_{\rm real} (r , p_r = 0 
,\mbox{\boldmath$L$}^2 , \mbox{\boldmath$L$} \cdot \mbox{\boldmath$S$}_{\rm 
eff})}{\partial \, r} = 0 \, .
\ee
Because of the (possibly non-linear) spin-orbit coupling, i.e. the dependence of $H_{\rm 
real}$ on $\mbox{\boldmath$L$} \cdot \mbox{\boldmath$S$}_{\rm eff}$, the orbital plane 
of these ``spherical'' orbits is not fixed in space. But the radial coordinate $r$ being 
constant, these orbits trace a complicated path on a sphere (hence the name). These 
orbits are the analogs, in our two-body problem, and in the approximation 
(\ref{eq3.12}), of similar exact ``spherical'' orbits for the geodesic motion of test 
particles in a Kerr spacetime \cite{wilkins}. Their existence (under some approximation) 
in the two-body problem is interesting for the following reason. One expects most black 
hole binary sources of interest for the LIGO/VIRGO/GEO network to have had the time to 
relax, under radiation reaction, to circular orbits. When the two black holes will get 
closer, these circular orbits will adiabatically shrink until they come close enough for 
feeling the effect of the spin-orbit coupling (which varies $\propto \, r^{-3}$). In 
some intermediate domain where the spin-orbit coupling is significant, but couplings 
quadratic in the spins are still small, the initially circular orbit will evolve into an 
adiabatic sequence of ``spherical'' orbits of the type just discussed. [We are here adding 
by hand the effect of radiation reaction, treated as an adiabatic perturbation of the 
conservative dynamics discussed in this paper.] These considerations indicate that, in 
first approximation, the total amount of gravitational radiation emitted by coalescing 
spinning black holes will be determined by the binding energy of the Last Stable 
Spherical Orbit (LSSO), i.e. the last stable solution of Eqs.~(\ref{eq3.19}), which will 
satisfy
\be
\label{eq3.20}
\frac{\partial \, H_{\rm real}}{\partial \, r} \ (r , p_r = 0 ,\mbox{\boldmath$L$}^2 , 
\mbox{\boldmath$L$} \cdot \mbox{\boldmath$S$}_{\rm eff}) = 0 \, , \ \frac{\partial^2 \, 
H_{\rm real}}{\partial \, r^2} \ (r , p_r = 0 ,\mbox{\boldmath$L$}^2 , 
\mbox{\boldmath$L$} \cdot \mbox{\boldmath$S$}_{\rm eff}) = 0 \, .
\ee

Before studying the energetics of these LSSO's let us mention the existence of other 
approximate first integrals in the dynamics of binary spinning black holes. Let us keep 
all the terms non-linear in $\mbox{\boldmath$S$}_{\rm eff}$, i.e. the full expression of 
$H_{\rm real} (\mbox{\boldmath$x$} , \mbox{\boldmath$p$} , \mbox{\boldmath$S$}_1 , 
\mbox{\boldmath$S$}_2)$, but let us try to approximately decouple the orbital motion 
from the spin degrees of freedom by considering that the two spin vectors evolve 
adiabatically (i.e. slowly on the orbital time scale), through Eqs.~(\ref{eq3.3}), 
(\ref{eq3.4}). In this adiabatic-spin approximation, the orbital motion is described by 
the Hamilton-Jacobi equation (\ref{eq2.17}), with an adiabatically fixed effective 
metric $g_{\rm eff}^{\alpha\beta}$. With the definition (\ref{eq2.36new}) of the 
quadratic-in-momenta term $Q_4 (p)$, one can check that, in this approximation, there 
will exist a two-body analog of the Carter constant for geodesic motion in Kerr 
\cite{carter}. Indeed, we have constructed our deformed Kerr metric (\ref{eq2.33}) so as 
to respect its separability properties. Let us work in an (adiabatic) 
Boyer-Lindquist-type coordinate system $(t,r,\theta,\varphi)$, as in Eq.~(\ref{eq2.33}). 
We find that the separability of the effective Hamilton-Jacobi equation yields the 
following first integrals (of the effective Hamiltonian)
\be
\label{eq3.21}
p_t = - E_{\rm eff} \, , \ p_{\varphi} = L_z \, ,
\ee
\be
\label{eq3.22}
p_{\theta}^2 + \frac{(L_z - a \, E_{\rm eff} \sin^2 \theta)^2}{\sin^2 \theta} + \mu^2 \, 
a^2 \cos^2 \theta = {\cal K} \equiv {\cal Q} + (L_z - a \, E_{\rm eff})^2 \, ,
\ee
\be
\label{eq3.23}
p_{\theta}^2 + \cos^2 \theta \left[ \frac{L_z^2}{\sin^2 \theta} + a^2 (\mu^2 - E_{\rm 
eff}^2) \right] = {\cal Q} \equiv {\cal K} - (L_z - a \, E_{\rm eff})^2 \, .
\ee
The last two equations are equivalent to each other, but, depending on the context, one 
can be more convenient than the other. Let us note the connection of the first integrals 
(\ref{eq3.21})--(\ref{eq3.23}) with the above analysis of the first integrals of the 
Hamiltonian depending only on the combination $\mbox{\boldmath$L$} \cdot 
\mbox{\boldmath$S$}_{\rm eff}$. The conservation of $L_z$, Eq.~(\ref{eq3.21}), 
corresponds to the conservation of $\mbox{\boldmath$L$} \cdot \mbox{\boldmath$S$}_{\rm 
eff}$, Eq.~(\ref{eq3.16}), while the conservation of ${\cal K}$ or ${\cal Q}$ 
corresponds to the conservation of $\mbox{\boldmath$L$}^2$.
Indeed, if we neglect the terms $\propto \, a^2$ in (\ref{eq3.23}) we get
\be
\label{eq3.24}
{\cal Q} \simeq p_{\theta}^2 + L_z^2 \ \frac{1 - \sin^2 \theta}{\sin^2 \theta} = 
\mbox{\boldmath$L$}^2 - L_z^2 \, .
\ee
This suggests that, even beyond the adiabatic-spin approximation, the quantities, now 
defined in an arbitrary frame as
\be
\label{eq3.25}
L_z \equiv \mbox{\boldmath$L$} \cdot \mbox{\boldmath$s$} \, , \ {\cal Q} \equiv 
\mbox{\boldmath$L$}^2 - (\mbox{\boldmath$L$} \cdot \mbox{\boldmath$s$})^2 + a^2 \, 
(\mbox{\boldmath$n$} \cdot \mbox{\boldmath$s$})^2 \, (\mu^2 - E_{\rm eff}^2) \, ,
\ee
will be (as well as $\mbox{\boldmath$S$}_{\rm eff}^2$ and ${\cal K} \equiv {\cal Q} + 
(L_z - a \, E_{\rm eff})^2$) conserved to a good approximation. We are mentionning here 
these approximate conservation laws because they could be helpful in qualitatively 
understanding the full two-body dynamics.

\subsection{Spherical orbits and last stable spherical orbits}\label{ssec3.2}

We discussed above the existence of spherical orbits under the assumption (or the 
approximation) that $H_{\rm real}$ depend only on the ``spin-orbit'' combination 
$\mbox{\boldmath$L$} \cdot \mbox{\boldmath$S$}_{\rm eff}$ (as it does at the 
linear-in-spin level). More generally, we have seen that if we treat the evolution of 
the spins as being adiabatic, we have the (approximate) first integrals (\ref{eq3.25}). 
If we use (as a heuristic mean of studying the main features of the orbital dynamics) 
this adiabatic approximation, we can define a family of spherical orbits by drawing on 
the conservation of the quantities (\ref{eq3.25}). Indeed, inserting the definitions 
(\ref{eq3.25}) into Eq.~(\ref{eq2.17}) we get an equation controlling the radial motion:
\begin{eqnarray}
\label{eq3.26}
&\Delta_r \, p_r^2 + 2 \, (4 - 3 \nu) \, \nu \, (GM)^2 \, p_r^4 / \mu^2 = \nonumber 
\\
&\displaystyle \frac{1}{\Delta_t \, (r)} \ [(r^2 + a^2) \, E_{\rm eff} - a \, L_z]^2 - 
(\mu^2 \, r^2 + 
{\cal Q} + (L_z - a \, E_{\rm eff})^2 ) \, .
\end{eqnarray}
The right-hand-side of Eq.~(\ref{eq3.26}) defines (when, $a^2 = \mbox{\boldmath$S$}_{\rm 
eff}^2 / M^2$, $L_z$ and ${\cal Q}$ are considered as adiabatic constants) a radial 
potential whose local minima, in $r$, determine (adiabatic) spherical orbits. The last 
stable spherical orbit is obtained when this radial potential has an inflection point. 
More precisely let us define
\be
\label{eq3.27}
R (r , E_{\rm eff} , L_z , {\cal Q}) \equiv ((r^2 + a^2) \, E_{\rm eff} - a \, L_z)^2 - 
\Delta_t (r) (\mu^2 \, r^2 + {\cal Q} + (L_z - a \, E)^2 ) \, .
\ee
The spherical orbits are the solutions of
\be
\label{eq3.28}
R (r , E_{\rm eff} , L_z , {\cal Q}) = \frac{\partial}{\partial \, r} \ R (r , E_{\rm 
eff} , L_z , {\cal Q}) = 0 \, .
\ee
The solutions of Eq.~(\ref{eq3.28}) yield a two-parameter family of solutions, along 
which, for instance, $r$ and $E_{\rm eff}$ are functions of $L_z$ and ${\cal Q}$. The 
last stable spherical orbit (LSSO) along such a family of solutions must satisfy the 
three equations
\be
\label{eq3.29}
R (r , E_{\rm eff} , L_z , {\cal Q}) = \frac{\partial}{\partial \, r} \ R (r , E_{\rm 
eff} , L_z , {\cal Q}) = \frac{\partial^2}{\partial \, r^2} \ R (r , E_{\rm eff} , L_z , 
{\cal Q}) = 0 \, .
\ee
There is a one-parameter family of LSSO's. For instance, one can take as free parameter 
the dimensionless ratio ${\cal Q} / L_z^2$ which is a measure of the maximum angle 
between the orbital plane and the equatorial plane defined by $\mbox{\boldmath$S$}_{\rm 
eff}$. [Note that ${\cal Q} = 0$ for an orbit in the equatorial plane.] For each value 
of this angle, and for each value of the effective spin parameter $a$, there will be 
some LSSO, with particular values of $r$, $E_{\rm eff}$ and $L_z$.

To study the values of the (effective and real) binding energy, and of the orbital 
angular momentum along this one-parameter family of LSSO's, it is convenient to work 
with slightly different variables. Let us introduce
\be
\label{eq3.30}
\bar{L}_z \equiv L_z - a \, E_{\rm eff} \, , \ {\cal K} \equiv {\cal Q} + \bar{L}_z^2 \, 
.
\ee
Let us also work with the radial variable $u \equiv 1/r$ and denote
\be
\label{eq3.31}
\bar A (u) \equiv \frac{\Delta_t \, (r)}{r^2} \equiv P_3^1 \, [A_{\rm orb}^{\rm PN} (u) 
+ a^2 \, u^2] \, .
\ee
We have
\be
\label{eq3.32}
r^{-4} \, R(r) \equiv U (u) \equiv (E_{\rm eff} - a \, \bar{L}_z \, u^2)^2 - \bar A (u) 
(\mu^2 + {\cal K} \, u^2) \, .
\ee
The equation $U (u) = 0$ (i.e. $R(r) = 0$) is now solved as
\be
\label{eq3.33}
E_{\rm eff} = W_a (u , \bar{L}_z , {\cal K}) \equiv a \, \bar{L}_z \, u^2 + \sqrt{\bar A 
(u) 
(\mu^2 + {\cal K} \, u^2)} \, .
\ee
The two-parameter family of spherical orbits is now obtained (as functions of the 
parameters $\bar{L}_z$ and ${\cal K}$) by solving $\partial \, W / \partial \, u = 0$, 
while the one-parameter family of LSSO's is obtained by solving $\partial \, W / 
\partial \, u = \partial^2 \, W / \partial \, u^2 = 0$. The advantage of this 
formulation is that it exhibits in the simplest way the analogy with the effective 
radial potential discussed in \cite{BD99,DJS2} for the pure (3~PN) orbital motion 
(without spin), namely
\be
\label{eq3.34}
W_0 (u,L) \equiv \sqrt{A(u) (\mu^2 + L^2 \, u^2)} \, ,
\ee
with $A(u) \equiv P_3^1 \, [A_{\rm orb}^{\rm PN} (u)]$. Apart from the replacement $L^2 
\rightarrow {\cal K}$, the only two differences between the spinning case 
((\ref{eq3.33})) and the spinless one ((\ref{eq3.34})) is the addition of the spin-orbit 
energy term $+ \, a \, \bar{L}_z \, u^2$, and the additional $a^2 \, u^2$ term in the PN 
expansion of $\bar A (u)$. [Note that $\bar A (u) \ne A(u) + a^2 \, u^2$ because the 
Pad\'eeing is done after the addition of $a^2 \, u^2$.] We have chosen to parametrize 
$W_a (u)$ in terms of $\bar{L}_z$ and ${\cal K}$ because it simplifies very much its 
expression and thereby renders more transparent the new physics incorporated in our 
effective one-body approach. The fact that $\bar{L}_z$ depends both on $L_z$ and $E_{\rm 
eff}$ is not a problem for solving Eq.~(\ref{eq3.32}) for $E_{\rm eff}$. Indeed, we are 
discussing a continuous family of solutions and it is essentially indifferent to 
parametrize them in terms of $L_z$ or $\bar{L}_z$. We could have introduced another 
effective potential $W'_a (r , L_z , {\cal Q})$ by solving $R (r , E_{\rm eff} , L_z , 
{\cal Q})= 0$, with Eq.~(\ref{eq3.27}), which would be more complicated, but which would 
describe the same physics. [Note that $W'_a (r)$ would directly exhibit the correct fact 
that the spin-orbit energy, for given $L_z$, decreases like $r^{-3}$, while this fact is 
hidden in $W_a (u)$ which assumes that $\bar{L}_z = L_z - a \, E_{\rm eff}$ is given.]

\subsection{Binding energy of last stable spherical orbits}\label{ssec3.3}

To get a first idea of the physical consequences of our effective one-body description 
of coalescing spinning black holes we have numerically investigated the properties of 
the one-parameter family of LSSO's. The most important quantity we are interested in is 
the binding energy at the last stable spherical orbit because it is the prime quantity 
determining the detectability of the GW emitted during the inspiral. We recall that the 
real, two-body energy is related to the effective energy entering the equations of the 
previous subsection through
\be
\label{eq3.35}
E_{\rm real} = M \, \sqrt{1+2\nu \left( \frac{E_{\rm eff}}{\mu} - 1 \right)} \, .
\ee
We are mostly interested in the (dimensionless) binding energy per unit total mass, say
\be
\label{eq3.36}
e \equiv \frac{E_{\rm real} - M}{M} = \sqrt{1+2\nu \left( \frac{E_{\rm eff}}{\mu} - 1 
\right)} - 1 \, .
\ee
The value of $e$ at the LSSO depends on three dimensionless parameters
\be
\label{eq3.37}
\nu_4 \equiv 4 \, \frac{\mu}{M} \equiv 4 \nu \, , \ \widehat a \equiv \frac{a}{M} \equiv 
\frac{\vert \mbox{\boldmath$S$}_{\rm eff} \vert}{M^2} \, , \ \cos \theta_{{\rm LS}} 
\equiv \frac{\bar{L}_z}{\sqrt{\cal K}} \, .
\ee
Here, the parameter $\nu_4$ (renormalized so that $0 < \nu_4 \leq 1$) determines the 
effect of having comparable masses ($\nu_4 \simeq 1$) rather than a large mass hierarchy 
($\nu_4 \ll 1$). The dependence of $e^{\rm LSO}$ on $\nu_4$ in absence of spins was 
studied in \cite{BD99,DJS2}. It was found that the ratio $e^{\rm LSO} / \nu_4$ was 
essentially linear in $\nu_4$ (even for $\nu_4$ as large as 1, corresponding to the 
equal-mass case)
\be
\label{eq3.38}
e_{S_{\rm eff} \, = \, 0}^{\rm LSO} \simeq - 0.014298 \, \nu_4 \, (1 + c_1 \, \nu_4) \, .
\ee
Here, the numerical value $- 1.4298\% = \frac{1}{4} \left(\sqrt{\frac{8}{9}} - 1 
\right)$ is one fourth the specific binding LSO energy of a test particle in 
Schwarzschild. The numerical coefficient $c_1$ which condenses the effect of resummed PN 
interactions was found to have a value $c_1^{2 \, {\rm PN}} \simeq 0.048$ at 2~PN and 
$c_1^{3 \, {\rm PN}} (\omega_s = 0) \simeq 0.168$ at 3~PN, and for $\omega_s = 0$. [The 
dependence of $c_1^{3 \, {\rm PN}}$ on $\omega_s$ is also roughly linear: $c_1^{3 \, 
{\rm PN}} (\omega_s) \simeq 0.168 + 0.0126 \, \omega_s$, at least when $- \, 10 \, \laq \, 
\omega_s \, \laq \, 0$.]

We expect that the dependence on $\nu_4$ of the spin-dependent effects will also be 
roughly linear (after factorization of an overall factor $\nu_4$ which comes from 
expanding the square root in Eq.~(\ref{eq3.36})). In the following we shall generally 
consider (in our numerical investigations) the case $\nu_4 = 1$, and concentrate on the 
dependence on the other parameters.

Let us clarify the meaning of the parameters $\widehat a$ and $\cos \theta_{\rm LS}$ 
introduced in Eq.~(\ref{eq3.37}). The quantities $\mbox{\boldmath$S$}_{\rm eff}$, 
$\bar{L}_z$ 
and ${\cal K}$ entering these definitions are all supposed to be computed at the last 
stable spherical orbit of an adiabatic sequence of spherical orbits (in the sense 
discussed above). Physically, we have in mind the sequence of inspiralling orbits driven 
by radiation reaction. Technically, we define $e (\nu_4 , \widehat a , \cos \theta_{\rm 
LS})$ by solving the effective radial potential problem defined in the previous 
subsection. We are aware of the fact that we cannot really attach to $\cos \theta_{\rm 
LS}$ the meaning of being the cosinus between $\mbox{\boldmath$L$}$ and 
$\mbox{\boldmath$S$}_{\rm eff}$ (as the name would suggest), but this is not important. 
What is important is that there is indeed a physical degree of freedom related to the 
misalignment between $\mbox{\boldmath$L$}$ and $\mbox{\boldmath$S$}_{\rm eff}$ at the 
LSSO and that we measure it by a parameter normalized so that $\cos \theta_{\rm LS} = 1$ 
(or $-1$) when all angular momenta are aligned (in this limit the concept of last stable 
circular equatorial orbit is meaningful and coincides with the $\cos \theta_{\rm LS} = 
1$ (or $-1$) limit of our formal definitions).

Note that, with the definitions (\ref{eq3.37}) and the additional definition $\bar{\ell} 
\equiv ({\cal K})^{1/2} / GM \, \mu$, the effective radial potential (\ref{eq3.33}) 
yields (in dimensionless form, $\widehat u \equiv GM / r$)
\be
\label{eq3.39}
\frac{E_{\rm eff}}{\mu} = \widehat{W}_{\widehat a} (\widehat u , \cos \theta_{\rm LS} , 
\bar{\ell}) = \widehat a \, \cos \theta_{\rm LS} \, \bar{\ell} \, \widehat{u}^2 + 
\sqrt{\bar A (\widehat u , \widehat{a}^2) (1 + \bar{\ell}^2 \, \widehat{u}^2 )} \, .
\ee
This form makes it clear that $E_{\rm eff}^{\rm LSSO} / \mu$, and therefore $e^{\rm 
LSSO}$, Eq.~(\ref{eq3.36}), will depend primarily on the combination (``projected value 
of $\widehat a$'')
\be
\label{eq3.40}
\widehat{a}_p \equiv \widehat a \, \cos \theta_{\rm LS} \simeq \frac{\mbox{\boldmath$k$} 
\cdot \mbox{\boldmath$S$}_{\rm eff}}{M^2}
\ee
where $\mbox{\boldmath$k$} = \mbox{\boldmath$L$} / \vert \mbox{\boldmath$L$} \vert$ (at 
the LSSO). For smallish spins, the combination $\widehat{a}_p$ is the only one entering 
the problem {\em linearly}. As recalled by the notation in (\ref{eq3.39}) the 
non-projected value of $\widehat a$ enters only quadratically in $\bar A (\widehat u , 
\widehat{a}^2)$.

Let us consider more closely the crucial quantity
\begin{eqnarray}
\label{eq3.41}
\widehat{a}_p &= &\frac{1}{(m_1 + m_2)^2} \left[ \left( 1 + \frac{3}{4} \, 
\frac{m_2}{m_1} \right) \, \mbox{\boldmath$k$} \cdot \mbox{\boldmath$S$}_1 + \left( 1 + 
\frac{3}{4} \, \frac{m_1}{m_2} \right) \, \mbox{\boldmath$k$} \cdot 
\mbox{\boldmath$S$}_2 \right] \nonumber \\
&= &\left( X_1^2 + \frac{3}{4} \, X_1 \, X_2 \right) \, \mbox{\boldmath$k$} \cdot 
\widehat{\mbox{\boldmath$a$}}_1 + \left( X_2^2 + \frac{3}{4} \, X_1 \, X_2 \right) \, 
\mbox{\boldmath$k$} \cdot \widehat{\mbox{\boldmath$a$}}_2 \, . 
\end{eqnarray}
In the second form, we have defined $X_1 \equiv m_1 / M$, $X_2 \equiv m_2 / M$ ($X_1 + 
X_2 = 1$, $X_1 \, X_2 = \nu$) and $\widehat{\mbox{\boldmath$a$}}_1 \equiv 
\mbox{\boldmath$S$}_1 / m_1^2$, $\widehat{\mbox{\boldmath$a$}}_2 \equiv 
\mbox{\boldmath$S$}_2 / m_2^2$. [We recall that a maximally spinning hole would have 
$\vert \widehat{\mbox{\boldmath$a$}}_1 \vert = 1$.]

An important question (for the relevance of the present work) is: what are the plausible 
values of $\widehat{a}_p$ in the sources that will be detected by LIGO/VIRGO/GEO?
Present astrophysical ideas about the formation of binary black holes 
\cite{postnov,PZ99} suggest neither that the holes be typically maximally spinning, nor 
that there be any correlation between the spin and angular momenta, i.e. between the 
directions of $\mbox{\boldmath$k$}$, $\mbox{\boldmath$S$}_1$ and 
$\mbox{\boldmath$S$}_2$. Not much is known either about the probable value of the mass 
ratio. To have an idea of the plausible values of $\widehat{a}_p$ (which is an algebraic 
quantity which can take positive or negative values) let us consider the random mean 
square (rms) value of $\widehat{a}_p$ under the assumption of random, uncorrelated 
directions $\mbox{\boldmath$k$}$, $\mbox{\boldmath$S$}_1$ and $\mbox{\boldmath$S$}_2$ 
(so that $\langle \widehat{a}_p \rangle = 0$). Let us assume (for simplicity) that $m_1 
= m_2$, i.e. $\nu_4 = 1$, which is the most favourable case because $e^{\rm LSSO} 
\propto \nu_4$. We assume also that $\langle \widehat{\mbox{\boldmath$a$}}_1^2 \rangle = 
\langle \widehat{\mbox{\boldmath$a$}}_2^2 \rangle \equiv (a_1^{\rm rms})^2$ is some 
given quantity (to be determined by astrophysical models). This yields for 
$\widehat{a}_p^{\rm rms} \equiv \sqrt{\langle \widehat{a}_p^2 \rangle}$
\be
\label{eq3.42}
\widehat{a}_p^{\rm rms} = \frac{7}{16} \, \frac{\sqrt 2}{\sqrt 3} \, a_1^{\rm rms} = 
0.357 \, a_1^{\rm rms} \, .
\ee
Even if $\widehat{a}_1^{\rm rms} = 1$ (which would mean that all black holes are 
maximally spinning) we get $\widehat{a}_p^{\rm rms} = 0.357$. However, we find it highly 
plausible that $\widehat{a}_1^{\rm rms}$ will be significantly smaller than 1. For 
instance, 
if we 
optimistically assume a uniform distribution of spin kinetic energy between 0 and the 
maximal value we would get $\widehat{a}_1^{\rm rms} = \frac{1}{\sqrt 2}$ and therefore 
$\widehat{a}_p^{\rm rms} = 7 / (16 \, \sqrt 3) = 0.253$. In view of these arguments, we 
find plausible that most LIGO/VIRGO binary black hole sources will have $\vert 
\widehat{a}_p \vert \, \laq \, 0.3$. This consideration is important because we shall 
see later that for such smallish values of $\widehat{a}_p$ the simple analytical approach 
advocated here seems to be quite reliable. However, one should also be able to compute
physically reliable (or, at least, sufficiently flexible) templates for fast
spinning binary black holes. As we said in the Introduction, we think that the EOB
approach can be an essential tool for this purpose, in conjunction with numerical
data, by feeding  the (necessarily sparse) numerical data into some
multi-parameter version of the EOB Hamiltonian.

This statistical estimate of the plausible value of $a_p$ suggests that a typical value 
of $\cos \theta_{\rm LS} \simeq \mbox{\boldmath$k$} \cdot \mbox{\boldmath$S$}_{\rm eff} 
/ \vert \mbox{\boldmath$S$}_{\rm eff} \vert$ is around $\pm \, 1 / \sqrt 3$. In our 
numerical 
estimates of $e^{\rm LSSO}$ we have used this value, as well as the (implausible) value 
$\cos \theta_{\rm LS} = \pm \, 1$ corresponding to perfect alignment.
As a first step towards exploring the ``parametric flexibility'' of the EOB
approach, we have studied the dependence of $e^{\rm LSSO}$ on the value of the
parameter $\omega_s$, as it appears in Eq.~(\ref{eq2.21}).
  We have done numerical simulations for three fiducial values: $\omega_s = 0 
\equiv \omega^{\rm DJS}$ 
\cite{DJSd}, $\omega_s = - 1987 / 840 \equiv \omega_s^{\rm BF}$ 
\cite{BF1,BF3}\footnote{This value corresponds to taking $\lambda 
= 0$  
in $\omega_s = - \, 11 \lambda / 3 - 1987/840$. Here, $\lambda$ denotes the natural 
ambiguity parameter entering the Blanchet-Faye
 framework. Note that the authors of Refs.~\cite{BF1,BF2,BF3} do not claim 
that $\lambda = 0$ is a preferred value. However, as $\lambda$ is expected to be of order 
unity we use $\lambda = 0$, i.e. $\omega_s = - \, 1987/840$ as a fiducial deviation from 
$\omega_s = 0$.}, and also 
for $\omega_s = - \frac{1}{2} \left( \frac{94}{3} - \frac{41 \, \pi^2}{32} \right) \equiv 
\omega_s^*$. Note the numerical values:
$\omega_s^{\rm BF} \simeq - 2.3655$, $\omega_s^* \simeq - 9.3439$.
The original motivation (when writing this paper, before the completion of the 
work \cite{DJSd} which determined the correct value of $\omega_s$) was to study the
sensitivity of our results to the ``3PN ambiguity''. We kept it here as an 
interesting case study of the sensitivity of EOB results to modifications
of the various coefficients entering the EOB Hamiltonian. The change from 
$\omega_s = 0$ to $\omega_s = \omega_s^{\rm BF}$ corresponds to a change of
the coefficient $a_4 \equiv a_4(\nu) / \nu$ from $18.688$ to $13.957$, i.e.
a fractional change of $ - 25.32 \%$. The 
value $\omega_s = \omega_s^*$ has the effect of completely cancelling the 3~PN 
contribution to the radial functions $A(u)$ and $\bar A (u)$. Therefore, choosing 
$\omega_s 
= \omega_s^*$ gives for the LSSO quantities the same results as the 2~PN 
effective-one-body Hamiltonian \cite{BD99}. Its consideration is useful for
 exhibiting the difference between the 2~PN-based results and the 3~PN-based 
ones. Our results are displayed in Table~\ref{Tab2} and Fig.~1.

\begin{table}
\begin{center}
\begin{tabular}{cc}
$\cos \theta_{\rm LS} = \pm \, 1 / \sqrt 3$ &$\cos \theta_{\rm LS} = \pm \, 1$ \\ 
\hline
\begin{tabular}{ccccccc}
$\widehat{a}_p$ & \qquad \quad &$\widehat a$ & \qquad \quad &$e^{\rm LSSO}$ & \qquad \quad 
&$\widehat{r}^{\rm LSSO}$ \\
$-0.6$ & \qquad \quad &$-1.039$ & \qquad \quad &$-0.01319$ & \quad \quad &$6.298$ \\
$-0.5$ & \qquad \quad &$-0.8660$ & \qquad \quad &$-0.01337$ & \quad \quad &$6.220$ \\
$-0.4$ & \qquad \quad &$-0.6928$ & \qquad \quad &$-0.01364$ & \quad \quad &$6.109$ \\
$-0.3$ & \qquad \quad &$-0.5196$ & \qquad \quad &$-0.01405$ & \quad \quad &$5.940$ \\
$-0.2$ & \qquad \quad &$-0.3464$ & \qquad \quad &$-0.01463$ & \quad \quad &$5.700$ \\
$-0.1$ & \qquad \quad &$-0.1732$ & \qquad \quad &$-0.01547$ & \quad \quad &$5.377$ \\
$0.$ & \qquad \quad &$0.$ & \qquad \quad &$-0.01670$ & \quad \quad &$4.954$ \\
$+0.1$ & \qquad \quad &$+0.1732$ & \qquad \quad &$-0.01859$ & \quad \quad &$4.391$ \\
$+0.2$ & \qquad \quad &$+0.3464$ & \qquad \quad &$-0.02203$ & \quad \quad &$3.580$ \\
$+0.3$ & \qquad \quad &$+0.5196$ & \qquad \quad &$-0.05146$ & \quad \quad &$1.344$ \\
$+0.4$ & \qquad \quad &$+0.6928$ & \qquad \quad &$-0.1790$ & \quad \quad &$0.7752$ \\
$ \ $ & \qquad \quad &$ \ $ & \qquad \quad &$ \ $ & \quad \quad &$ \ $ \\
$ \ $ & \qquad \quad &$ \ $ & \qquad \quad &$ \ $ & \quad \quad &$ \ $
\end{tabular}
&\begin{tabular}{ccccccc}
$\widehat{a}_p$ & \qquad \quad &$\widehat a$ & \qquad \quad &$e^{\rm LSSO}$ & \qquad \quad 
&$\widehat{r}^{\rm LSSO}$ \\
$-0.6$ & \qquad \quad &$-0.6$ & \qquad \quad &$-0.01150$ & \qquad \quad &$7.344$ \\
$-0.5$ & \qquad \quad &$-0.5$ & \qquad \quad &$-0.01207$ & \quad \quad &$6.989$ \\
$-0.4$ & \qquad \quad &$-0.4$ & \qquad \quad &$-0.01271$ & \quad \quad &$6.623$ \\
$-0.3$ & \qquad \quad &$-0.3$ & \qquad \quad &$-0.01345$ & \quad \quad &$6.242$ \\
$-0.2$ & \qquad \quad &$-0.2$ & \qquad \quad &$-0.01433$ & \quad \quad &$5.841$ \\
$-0.1$ & \qquad \quad &$-0.1$ & \qquad \quad &$-0.01538$ & \quad \quad &$5.415$ \\
$0.$ & \qquad \quad &$0.$ & \qquad \quad &$-0.01670$ & \quad \quad &$4.954$ \\
$+0.1$ & \qquad \quad &$+0.1$ & \qquad \quad &$-0.01842$ & \quad \quad &$4.439$ \\
$+0.2$ & \qquad \quad &$+0.2$ & \qquad \quad &$-0.02091$ & \quad \quad &$3.833$ \\
$+0.3$ & \qquad \quad &$+0.3$ & \qquad \quad &$-0.02529$ & \quad \quad &$3.005$ \\
$+0.4$ & \qquad \quad &$+0.4$ & \qquad \quad &$-0.04930$ & \quad \quad &$1.538$ \\
$+0.5$ & \qquad \quad &$+0.5$ & \qquad \quad &$-0.1048$ & \quad \quad &$1.194$ \\
$+0.6$ & \qquad \quad &$+0.6$ & \qquad \quad &$-0.1474$ & \quad \quad &$0.9792$
\end{tabular}
\end{tabular}
\medskip
\caption{Binding energies $e \equiv ({\cal E}_{\rm real} / M) - 1$ and (effective 
Boyer-Lindquist) radii of last stable spherical orbits (LSSO) for equal-mass spinning 
binary systems. The LSSO's depend on two independent parameters: 
$\vert \widehat{\mbox{\boldmath$a$}} \vert \equiv \vert \mbox{\boldmath$S$}_{\rm eff} 
\vert / M^2$ and $\cos 
\theta_{\rm LS} \equiv \bar{L}_z / \sqrt{{\cal K}}$ (which is, morally, the cosine of the 
angle between the orbital angular momentum and the effective spin). The combined parameter 
$\widehat{a}_p \equiv \vert \widehat{\mbox{\boldmath$a$}} \vert \, \cos \theta_{\rm LS}$ 
(projected spin) plays a primary 
role for moderate spins. The algebraic quantity $\widehat{a}$ is defined as $+ \, \vert 
\widehat{\mbox{\boldmath$a$}} \vert$ if $\cos \theta_{\rm LS} > 0$ and $- \, \vert 
\widehat{\mbox{\boldmath$a$}} \vert$ if $\cos \theta_{\rm LS} < 0$. All quantities are 
computed from the 3~PN-level Pad\'e-resummed effective-one-body Hamiltonian (\ref{eq2.44}) 
with $\omega_s = 0$.}
\label{Tab2}
\end{center}
\end{table}

\begin{figure}
\begin{center}
\hspace{-0.8cm} 
\epsfig{file=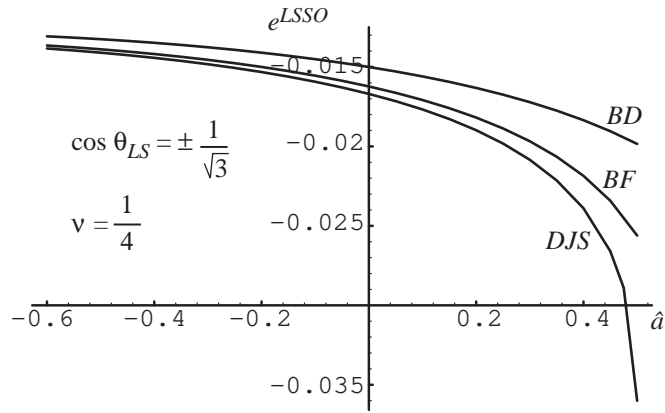} 
\medskip
\caption{\sl Dependence of the binding energy $e \equiv ({\cal E}_{\rm real} / M) - 1$ of
the LSSO on the effective spin parameter $\widehat a \equiv S_{\rm eff} / M^2$ (taken with
the sign of $\cos \theta_{\rm LS}$). We consider an equal-mass system with a typical
misalignment angle $\cos \theta_{\rm LS} = \pm \, 1 / \sqrt 3$. All three curves used the
Pad\'e-resummed Effective-one-body approach. The lower curves use a 3~PN-level
Hamiltonian: the lowest one uses $\omega_s = 0 = \omega_s^{\rm DJS}$ $[10]$, while
the middle one uses $\omega_s = - 2.365 = \omega_s^{\rm BF}$ $[17]$. The upper curve
uses $\omega_s = - 9.344 = \omega_s^*$, which is (essentially) equivalent to using a 
2~PN-level Hamiltonian $[8]$.}
\label{Fig1}
\end{center}
\end{figure}

The most important conclusion we wish to draw from these results is that, when $\widehat a 
\, \laq \, 0.3$ (which, as we argued above, covers a large domain of the physically 
relevant cases), the binding energy at the LSSO seems to be reliably describable by the 
simple analytical EOB Hamiltonian defined above.
 Indeed, the differences between: (i) the non-spinning 
case and the spinning ones, and (ii) the 2~PN orbital approximation and the 3~PN one, are 
all 
quite moderate (which indicates that the effective one-body approach is effective in 
resumming PN interactions near the LSO). Furthermore, the difference 
between: 
(iii) the spinning 3~PN case with $\omega_s = 0 = \omega_s^{\rm DJS}$ \cite{DJS2}, and the 
same case with 
$\omega_s = \omega_s^{\rm BF}$ \cite{BF3} is rather small. This is a testimony of
the robustness of the EOB approach. A change of its 3PN coefficient by $25 \%$ does
not affect much the physical predictions. This robustness at the 3PN level is
indicative of some robustness against the addition of higher PN effects.
Note also 
(from table~\ref{Tab2}) the confirmation that when $\vert \widehat{a}_p \vert \, \laq \, 
0.2$, the binding energy at the LSSO depends nearly only on the projected effective spin 
parameter $\widehat{a}_p = \widehat{a} \cos \theta_{\rm LS}$, with a very weak dependence 
on the value of $\cos \theta_{\rm LS}$.

On the other hand, it must be admitted that when, say, $\widehat{a} \, \gaq \, 0.4$ the 
differences between the three cases (i), (ii), (iii) become so large, and the radius of 
the 
LSSO becomes so small, that the present spin-dependent
 EOB predictions cannot be quantitatively trusted. 
[However, 
as discussed in more detail below, we think that they remain qualitatively correct.] If 
the 
orbital dynamics were well 
described by the 2~PN-level orbital EOB metric (i.e. if $\omega_s$ had been near $-9$; see 
upper curve in Fig.~\ref{Fig1}), the 
binding energy, even in such extreme cases, would differ only moderately from the 
non-spinning case, and we could trust the EOB-plus-spin predictions. However, as 
$\omega_s$ is zero the 3~PN EOB $+$ spin predictions become, for $\cos 
\theta_{\rm LS} \simeq 1$ and $a \, \gaq \, 0.4$, very different from the 2~PN ones and 
quite sensitive to the numerical values of the expansion coefficients entering the
EOB potentials. Let us note, however, that in all cases (even the most extremely 
spinning ones) if the spin parameter $\widehat{a}_p$ is {\em negative} (i.e. if the 
effective spin vector, whatever be its magnitude, has a negative projection on the 
orbital angular momentum) the EOB predictions become extremely reliable because all the 
differences between the cases (i), (ii), (iii) become quite small.

All these results are easy to interpret physically. This can be seen from the basic 
equations of the EOB approach which simplify so much the description of the physical 
interactions by representing them as slightly deformed versions of the well-known 
gravitational physics of test particles in Schwarzschild or Kerr geometries. Indeed, the 
basic equation of the EOB approach determining the binding of the LSO is 
Eq.~(\ref{eq3.39}) which differs from its well-known\footnote{Actually, as far as we 
know, the Kerr limit of Eq.~(\ref{eq3.39}) has never been written down before. Usual 
treatments \cite{wilkins} use the more complicated effective radial potential $W'_a (r , 
L_z , {\cal Q})$. Anyway, the physics is the same, but it is more cleanly presented in 
(\ref{eq3.39}).} Kerr limit (i.e. $\nu \rightarrow 0$) only by the change
\be
\label{eq3.43}
A_K (\widehat u , \widehat{a}^2) = 1 - 2 \, \widehat u + \widehat{a}^2 \, \widehat{u}^2 
\rightarrow \bar A (\widehat u , \widehat{a}^2) = P_3^1 \, [1 - 2 \widehat u + 
\widehat{a}^2 \, \widehat{u}^2 + 2 \nu \, \widehat{u}^3 + a_4 (\nu) \, \widehat{u}^4 ] \, 
.
\ee
The crucial point (which is, finally, the most important new information obtained by the 
2~PN and 3~PN orbital calculations) is that the 2~PN and 3~PN additional terms to the 
radial function $A_{a \, = \, 0}^{\rm PN} (\widehat{u})$ have both {\em positive} 
coefficients. 
This means that, even before the addition of the effect of spin (which leads to a $+ \, 
\widehat{a}^2 \, \widehat{u}^2$ additional term in $A_K (\widehat{u})$, corresponding to 
the famous $+ \, a^2$ term in $\Delta_K (r) = r^2 - 2 Mr + a^2$) the main effect of 
non-linear orbital interactions for comparable masses is {\em ``repulsive''}, i.e. 
correspond 
to a partial screening of the basic Schwarzschild attractive term $1 - 2 \, \widehat{u} 
= 1 - 2 \, GM / c^2 \, r$ by the addition of repulsive terms $\propto + \, \nu / 
r^3$ and $+ \, \nu / r^4$. Now, paradoxically, the addition of a repulsive term leads to 
a more tightly bound LSSO because the {\em less attractive}, but still 
attractive\footnote{Remember that we Pad\'e resum $A (\widehat u)$ and $\bar A (\widehat 
u)$ to ensure that these functions qualitatively behave like $1 - 2 \, \widehat{u}$ or 
$1 - 2 \, \widehat{u} + \widehat{a}^2 \, \widehat{u}^2$ (for $\widehat{a}^2 < 1$), i.e. 
(generically) have a simple zero near $\widehat{r} = \widehat{u}^{-1} = 1 + \sqrt{1 - 
\widehat{a}^2}$, 
which means that the effective metric becomes ``infinitely attractive'' at some 
deformed horizon.}, radial function $A (\widehat{u} , \widehat{a}^2 = 0)$ will be able to 
``hold'' a particle in spherical orbit down to a {\em lower} orbit. In other words, when 
a radial potential becomes less attractive, its LSSO gets closer to the horizon, and the 
binding energy of the LSSO becomes more negative. This being said, one understands 
immediately the additional effects due to the spin interaction. There are basically two 
such effects: (a) a linear ``spin-orbit'' effect linked to the $+ \, \widehat{a}_p \, 
\bar \ell \, \widehat{u}^2$ term in (\ref{eq3.39}) (with $\widehat{a}_p \equiv \widehat{a} 
\cos 
\theta_{\rm LS}$), and (b) a non-linear spin-quadratic modification of the metric 
coefficient, i.e. the additional $+ \, \widehat{a}^2 \, \widehat{u}^2$ term in $\bar A 
(\widehat{u} , \widehat{a}^2)$ (or in $A_K (\widehat{u}) = 1 - 2 \, \widehat{u} + 
\widehat{a}^2 \, \widehat{u}^2$). The crucial points are that: (1) when $\widehat{a}_p < 
0$, i.e. $\cos \theta_{\rm LS} < 0$ (coarse antialignment of angular momenta) the 
dominant linear spin-orbit coupling is {\em attractive} and therefore pushes the LSSO 
{\em upwards}, towards a less bound orbit, while, (2) when $\widehat{a}_p > 0$, i.e. 
$\cos \theta_{\rm LS} > 0$ (coarse alignment of angular momenta) {\em both} the linear 
spin-orbit coupling $+ \, \widehat{a}_p \, \bar \ell \, \widehat{u}^2$ and the 
spin-quadratic 
additional term $+ \, \widehat{a}^2 \, \widehat{u}^2$ are {\em repulsive} and tend to draw 
the 
LSSO {\em downwards}, i.e. closer to the horizon, in a more bound orbit. Therefore we 
see that, when $\widehat{a}_p >0$, all the new effects (the $\nu$-dependent non-linear 
orbital interactions and the spin effects) tend in the same direction: towards a closer, 
more bound orbit. As the existence of a LSSO is due to a delicate balance between the 
attractive gravitational effects and the usual repulsive (``centrifugal'') effect of the 
orbital angular momentum (i.e. the term $+ \, \bar{\ell}^2 \, \widehat{u}^2 \propto + \, 
L^2 / r^2$ in (\ref{eq3.39})), when several attractive effects combine their action, 
they start having a large effect on the binding of the LSSO. This is well-known to be the 
case for circular, equatorial, corotating $(\widehat{a}_p = + \, \widehat{a})$ orbits of a 
test particle in Kerr, which feature, in the case of an extreme Kerr $(\widehat{a} = 1)$ 
an LSO at $\widehat r = 1$, with $\mu$-fractional binding $(E_{\rm eff} - \mu) / \mu = 1 
/ \sqrt 3 - 1 = - 0.42265$ (corresponding to $e \simeq \nu \, (E_{\rm eff} - \mu) / \mu 
\simeq - 0.10566 \, \nu_4$). It is also well-known that, again for extreme Kerr, a 
counterrotating $(\widehat{a}_p = - \, \widehat{a})$ circular, equatorial orbit in extreme 
Kerr has an LSO at $\widehat r = 9$, with $\mu$-fractional binding $(E_{\rm eff} - \mu) 
/ \mu = 5 / (3 \, \sqrt 3 ) - 1 = - 0.037750$ (corresponding to $e \simeq - 0.0094374 \, 
\nu_4$). What is less well-known is that the extreme binding of the circular, 
equatorial, corotating LSO around an extreme Kerr is not representative of the 
binding of typical LSSO's around typical (or even extreme) Kerr holes. Indeed, when 
$\cos \theta_{\rm LS} \ne \pm \, 1$ (i.e. in more invariant language, when ${\cal Q} \ne 
0$) 
and when $\widehat{a} < 1$, the LSSO is, in general, moderately perturbed away from the 
Schwarzschild value $\widehat{r}_{\rm LSO} = 6$ and its binding is correspondingly 
moderately different from its Schwarzschild limit $e_{\rm Schw}^{\rm LSO} \simeq - 
0.014298 \, \nu_4$. The present work has shown that the location of the LSSO's for 
binary spinning holes can be rather simply obtained, in the EOB approach, by balancing 
in the specific way of Eq.~(\ref{eq3.39}) the centrifugal effect of the orbital angular 
momentum against the overall attractive effect of gravity, but with the critical 
addition of the 2~PN and 3~PN repulsive terms, of the spin-quadratic repulsive term, and 
of the indefinite-spin effect of the spin-orbit interaction.

\subsection{Expected spin of the hole formed by the coalescence of two spinning 
holes}\label{ssec3.4}

The last topic we wish to discuss concerns the expected result of the coalescence of two 
holes. In particular, we are interested in estimating the maximal spin that the final 
hole, resulting from the coalescence of two spinning holes, might have. It was estimated 
in \cite{BD00,BD01} (by using the EOB approach) that the coalescence of two 
non-spinning holes of the same mass $m_1 = m_2 = M/2$ leads (after taking into account 
the effect of gravitational radiation on the orbital evolution and on the loss of energy 
and angular momentum) to the formation of a rotating black hole of mass $M_{\rm BH} 
\simeq (1 - \varepsilon_{\rm rd}) \, 0.976 \, M$ and spin parameter $\widehat{a}_{\rm 
BH} \simeq 0.80$ [we have included a factor $1 - \varepsilon_{\rm rd}$ in $M_{\rm BH}$ 
to take into account the energy loss during the ring-down. Ref.~\cite{BD01} found 
$\varepsilon_{\rm rd} \simeq 0.7\%$]. The fractional energy $0.976 - 1 = - 0.024$ 
roughly corresponds to the (adiabatically estimated) LSO binding energy ($- 0.015$ in 
the 2~PN-based estimate of \cite{BD99}) minus the energy per unit mass radiated during 
the plunge ($\sim - 0.007$ \cite{BD01}). We shall leave to future work a similar 
estimate, for the 3~PN-plus-spin case, of the amount of energy emitted in GW. We wish 
here to focus on the issue of the spin of the final hole. The above value 
$\widehat{a}_{\rm BH} \simeq 0.80$ is rather close to the maximal value 
$\widehat{a}_{\rm max} = 1$ and there arises the question of whether an EOB treatment of 
the coalescence of two spinning holes might not formally predict a final value of 
$\widehat{a}_{\rm BH}$ larger than one! By ``EOB treatment'' we mean here a completed 
version of the EOB approach (as in \cite{BD00} at the 2~PN, non-spinning level) obtained 
by: (i) adding a resummed radiation force to the ``conservative'' EOB dynamics, and (ii) 
pushing the calculation of the EOB evolution down to its point of unreliability (near 
the last {\em unstable} orbit) where it is matched to a perturbed-single-black-hole 
description. A zeroth approximation to this completed EOB approach is the one we study 
in this paper: an adiabatic sequence of solutions of the conservative dynamics, 
terminated at the LSO. In this approach one entirely neglects the losses of energy and 
angular momentum during the plunge phase following the crossing of the LSO. The numbers 
recalled above show that the energy loss during the plunge (and the ring-down) is not 
negligible compared to the binding energy at the LSO. However, for the present question 
this is not a problem. What is important is that the angular momentum loss during plunge 
is a very small fraction (a percent or so) of the angular momentum at the LSO, and that 
the final mass of the black hole is nearly equal to $M = m_1 + m_2$. This leads us to 
the following zeroth order estimate of the spin parameter of the final hole:
\be
\label{eq3.44}
\widehat{a}_{\rm BH} \simeq \frac{ \vert \mbox{\boldmath$J$} \vert^{\rm LSSO}}{(E_{\rm 
real}^{\rm LSSO})^2} \simeq \frac{ \vert \mbox{\boldmath$J$} \vert^{\rm LSSO}}{M^2} \, .
\ee
In view of the exact conservation of $\mbox{\boldmath$J$}$ in our conservative EOB 
(real) dynamics, it is clear that it is $\vert \mbox{\boldmath$J$} \vert^{\rm LSSO}$ 
which is a good measure of the total angular momentum of the final spacetime, i.e. of 
the final black hole.

\begin{figure}
\begin{center}
\hspace{-0.8cm} 
\epsfig{file=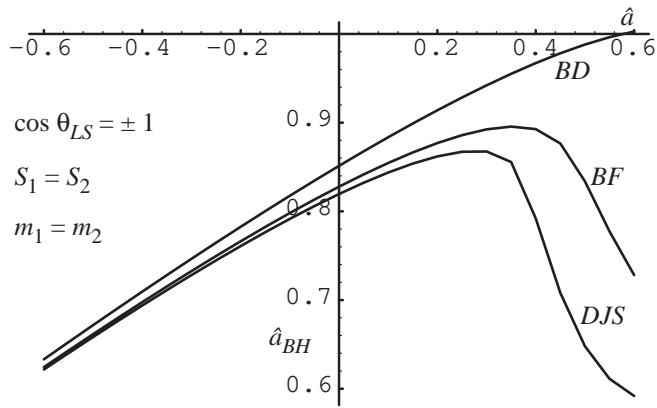}
\medskip
\caption{\sl Approximate prediction for the spin parameter $\widehat{a}_{\rm BH} \simeq 
\vert \mbox{\boldmath$J$} \vert^{\rm LSSO} / M^2$ of the black hole formed by the 
coalescence of two identical spinning holes (with spins parallel or antiparallel to the 
orbital angular momentum). The horizontal axis is the effective spin parameter $\widehat a 
= \frac{7}{8} \, \widehat{a}_1 = \frac{7}{8} \, \widehat{a}_2$. The three curves 
correspond 
to the three cases plotted in Fig.~\ref{Fig1}. Note the prediction (robust under
changing the 3~PN contribution to the effective potential by 25 \%) 
that the final spin parameter is always sub-extremal, and reaches a maximum 
$\widehat{a}_{\rm BH} \simeq 0.87$ for $\widehat a \simeq + \, 0.3$.}
\label{Fig2}
\end{center}
\end{figure}

We are facing here a potential consistency problem of this simple-minded EOB treatment: 
when computing (\ref{eq3.44}) for spinning configurations does one always get 
$\widehat{a}_{\rm BH} < 1$? One might worry that, starting with a value of 
$\widehat{a}_{\rm BH} \simeq 0.80$ for non-spinning holes, the addition of large spins 
on the holes might quickly exceed the extremal limit. It is plausible that the most 
dangerous situation is the ``aligned case'', where all the angular momenta, 
$\mbox{\boldmath$L$}$, $\mbox{\boldmath$S$}_1$ and $\mbox{\boldmath$S$}_2$ are parallel 
(or antiparallel). In this case the numerator of Eq.~(\ref{eq3.44}) reads
\be
\label{eq3.45}
J^{\rm LSSO} = L_z^{\rm LSSO} + S_1 + S_2 \, ,
\ee
while the spin parameter of the effective metric reads
\be
\label{eq3.46}
\widehat a = \widehat{a}_p = \frac{\mbox{\boldmath$k$} \cdot \mbox{\boldmath$S$}_{\rm 
eff}}{M^2} = \left( X_1^2 + \frac{3}{4} \, \nu \right) \, \widehat{a}_1 + \left( X_2^2 + 
\frac{3}{4} \, \nu \right) \, \widehat{a}_2 \, .
\ee
Here, we consider $S_1$, $S_2$ and $\widehat{a}_1 \equiv S_1 / m_1^2$, $\widehat{a}_2 
\equiv S_2 / m_2^2$ as algebraic numbers (positive or negative). This allows us to 
investigate also the case where the spins might be antiparallel to $\mbox{\boldmath$k$}$. 
For simplicity, we shall only study the symmetric case where $m_1 = m_2$ and $S_1 = 
S_2$. For this case
\be
\label{eq3.47}
\widehat a = \widehat{a}_p = \frac{7}{8} \, \widehat{a}_1 \, ,
\ee
and
\be
\label{eq3.48}
\widehat{a}_{\rm BH} = \frac{J^{\rm LSSO}}{M^2} = \frac{1}{4} \, \widehat{L}_z^{\rm 
LSSO} + \frac{1}{2} \, \widehat{a}_1 = \frac{1}{4} \, \widehat{L}_z^{\rm LSSO} + 
\frac{4}{7} \, \widehat{a}_p
\ee
where the dimensionless orbital angular momentum $\widehat{L}_z \equiv L_z / \mu \, M$ 
is related to the dimensionless quantity (when $\cos \theta_{\rm LS} = 1$) $\bar{\ell} 
\equiv \sqrt{\cal K} / \mu \, M = \bar{L}_z / \mu \, M$ appearing in (\ref{eq3.39}) 
through
\be
\label{eq3.49}
\widehat{L}_z = \bar{\ell} + \widehat{a} \, \frac{E_{\rm eff}}{\mu} \, .
\ee
It is interesting to note that, even in the case where both holes are extreme 
($\widehat{a}_1 = \widehat{a}_2 = 1)$ the maximum value of the effective spin parameter 
is $\widehat{a}_{\rm max} = \frac{7}{8} < 1$. We have numerically investigated the 
quantity $\widehat{a}_{\rm BH}$, Eq.~(\ref{eq3.48}), as a function of the effective 
$\widehat{a} = \widehat{a}_p$. The result is plotted in Fig.~\ref{Fig2} for different 
values of the 3~PN parameter $\omega_s$.

We see that the final spin parameter reaches a maximum for a positive value of 
$\widehat{a}_p$, i.e. for parallel (rather than antiparallel) spins. For the correct
3PN value $\omega_s = 0$
 the maximum value of $\widehat{a}_{\rm BH}$ is comfortably below 1: namely,
 $\widehat{a}_{\rm BH}^{\rm max} \simeq 0.87$, reached for $\widehat{a}_p 
\simeq + \, 0.3$. This is not much larger than the value $\widehat{a}_{\rm BH} \simeq 
0.82$ 
obtained for $\widehat{a}_p = 
0$. We find that this is a nice sign of the consistency of the EOB approach. This 
consistency was not a priori evident. In fact for $\omega_s \leq -9$ one gets a maximum 
value of $\widehat{a}_{\rm BH}$ slightly larger than 1. In particular, note that the 
2~PN treatment of the orbital dynamics (obtained for $\omega_s = \omega_s^* \simeq - 
9.3439$; upper curve in Fig.~\ref{Fig2}) formally leads to problematic over-extreme values 
of $\widehat{a}_{\rm BH}$. 
This may be interpreted as a confirmation of the need of ``repulsive'' 3~PN effects 
(i.e. $\omega_s + 9 \gg 1$). It is (a posteriori!) easy to understand physically why 
$\widehat{a}_{\rm BH}$, after reaching a maximum, then decreases when one adds more spin 
on the two black holes. Indeed, there is here a competition between two effects: adding 
spin on the holes (i.e. increasing $\widehat{a}_p$), on the one hand directly 
contributes to augmenting $\widehat{a}_{\rm BH}$ through the second term of the RHS of 
Eq.~(\ref{eq3.48}), but, on the other hand, indirectly contributes to reducing the total 
$J^{\rm LSSO}$ by reducing $\widehat{L}_z^{\rm LSSO}$ (indeed, as we explained above, 
positive spin leads to an LSO orbit closer to the horizon, and therefore with less 
orbital angular momentum). The first effect wins for smallish spins, while the second 
(more non-linear) effect wins for larger spins.


\section{Conclusions}\label{sec4}

We started by recalling the need of techniques for accelerating the convergence of the 
post-Newtonian (PN) expansions in the last stages of the inspiral of binary systems. We 
summarized the evidence (Table~\ref{Tab1}) showing the remarkable convergence properties 
of the best current resummation technique: the effective one-body (EOB) approach of 
Refs.~\cite{BD99,DJS2}. We showed how to generalize the EOB approach to the case of two 
spinning black holes with comparable masses $(\nu = \mu / M \sim 1/4)$.
 As a first step towards computing the spin-dependent EOB Hamiltonian 
 we constructed an 
effective metric, which can be viewed either as a $\nu$-deformation of the Kerr metric or 
as a spin-deformation of the $\nu$-deformed effective metric. The effective spin entering 
this deformed Kerr metric is $M \, 
\mbox{\boldmath$a$} \equiv \mbox{\boldmath$S$}_{\rm eff} \equiv \left( 1 + \frac{3}{4} \, 
\frac{m_2}{m_1} \right) \, \mbox{\boldmath$S$}_1 + \left( 1 + \frac{3}{4} \, 
\frac{m_1}{m_2} \right) \, \mbox{\boldmath$S$}_2$. The introduction of this effective 
$\mbox{\boldmath$a$}$ allows one to combine in a simple manner all (PN leading) 
spin-orbit coupling effects, and most of the spin-spin ones, with the rather complex but 
important 3~PN effects, which have been incorporated only recently in the EOB approach 
\cite{DJS2}. 

We have also constructed a more complicated modified effective Hamiltonian, 
Eq.~(\ref{eq2.56}), which separately depends on two (effective) spin vectors, $M 
\mbox{\boldmath$a$}_0 \equiv \mbox{\boldmath$S$}_0 \equiv \left(1 + \frac{m_2}{m_1} 
\right) \, 
\mbox{\boldmath$S$}_1 + \left(1 + \frac{m_1}{m_2} \right) \mbox{\boldmath$S$}_2$, 
and $\mbox{\boldmath$\sigma$} \equiv \mbox{\boldmath$S$}_{\rm eff} - \mbox{\boldmath$S$}_0 
\equiv - \frac{1}{4} \left( \frac{m_2}{m_1} \ \mbox{\boldmath$S$}_1 + \frac{m_1}{m_2} \ 
\mbox{\boldmath$S$}_2 \right)$, and which allows a (hopefully) more accurate 
representation 
of spin-spin effects. We recommend the simultaneous consideration of $H_{\rm eff}$ and 
$H'_{\rm eff}$ to determine the domain of trustability of the presently constructed
 spin-dependent EOB Hamiltonian. 
Namely, when $H_{\rm real} = \bar f (H_{\rm eff})$ and $H'_{\rm real} = \bar f (H'_{\rm 
eff})$ lead to numerically very similar evolutions, 
one is entitled to trusting them both; while a significant difference in their predictions 
signals a breakdown of the trustability of the simple EOB Hamiltonian proposed here.

The present paper has only investigated a few aspects of the physics predicted by our 
spin-generalized EOB approach. In particular, as a first cut toward understanding the 
relevance of our construction for gravitational wave (GW) observations we have discussed 
the approximate existence of ``spherical orbits'' (orbits with fixed radial coordinate, 
as in the Kerr metric) and we studied the binding energy of the last stable spherical 
orbits 
(LSSO). A message of this study is that, for most physically 
relevant cases (in the parameter space where one randomly varies all angles and all spin 
values), the results are only weakly dependent on the exact numerical values of the
3PN coefficients. Moreover, they exhibit moderate deviations from the 
non-spinning case (see Fig.~1 and Table~II). 
To give a numerical flavor of the effects of spin we note that, when the projected spin 
parameter $\widehat{a}_p = \mbox{\boldmath$k$} \cdot \mbox{\boldmath$a$} / M$, 
Eq.~(\ref{eq3.41}), is smaller than about $+ \, 0.2$, its effect on the fractional binding 
energy ($e \equiv (E_{\rm real} - M) / M$) of the LSSO is, approximately,
\be
\label{eq4.1}
100 \, e^{\rm LSSO} \simeq - 1.43 \, \nu_4 (1 + 0.168 \, \nu_4) - 0.806 \, \nu_4 (1 + 
0.888 
\, \nu_4) \, \widehat{a}_p \, ,
\ee
where $\nu_4 \equiv 4 \, \nu \equiv 4 \, m_1 \, m_2 / (m_1 + m_2)^2 \leq 1$. As in most 
cases (random angles, random spin-kinetic energies) it is plausible that $\vert 
\widehat{a}_p \vert \, \laq \, a_p^{\rm rms} \sim 0.25$, we expect that spin effects will 
only 
modify the energy emitted as gravitational waves up to the LSSO by less than about $0.6\% 
\, \nu_4 \, 
M$.

Such an increase, though modest, is still a significant fractional modification of the 
corresponding
energy loss predicted for non-spinning systems ($e_0 = - 1.67\% \, M$ for $\nu_4 = 1$). 
In fact, this effect might cause an important {\em bias} in the first observations. If 
the intrinsic spins of the holes can (at all) take large values, the highest 
signal-to-noise-ratio events in the first years of LIGO observations might select binary 
systems with rather large and rather aligned spins. It is therefore important to include 
spin effects in the data analysis of coalescing black holes. We have argued that, in most 
cases, the simple-minded 
generalized EOB approach presented here should be a reliable analytical tool 
for describing the dynamics of two spinning holes and for computing a catalogue of 
gravitational waveforms, to be used as matched filters in the detection of GW's. However, 
it must be admitted that, in 
the cases where the effective spin vector is coarsely (positively) aligned 
with the orbital angular momentum, and where the spins are so large that $\widehat a \, 
\gaq \, 0.4$, the predictions from the above-introduced
 EOB Hamiltonian start  predicting 
LSSO radii so near the ``effective horizon'' where $\Delta_t (r) = 0$,
 that they cannot be 
quantitatively relied upon. [Though I would still argue that they can be qualitatively 
trusted, in view of the simple physics they use; see subsection~\ref{ssec3.3}.] We give 
some examples of that in Table~\ref{Tab2}. In such cases the EOB 
approach does predict much larger energy losses, possibly larger than $10\% \, M$. In 
these 
cases, the uncertainty in the waveform may be so large that one may need 
the type of non-linear filtering search algorithm advocated in Ref.~\cite{FH98}. We wish, 
however, to emphasize the differences between our treatment and conclusions and those of 
Flanagan and Hughes. These authors defined the ``merger'' phase as (essentially) what 
comes after the binary system crosses the non-spinning LSO (around 6~$GM$), and they 
assumed that the signal from the ``merger'' phase can only be obtained from numerical 
relativity. Moreover, they optimistically assumed that (in all cases) $10\% \, M$ are 
emitted in GW energy during the merger phase, and 3\% during the subsequent ring-down 
phase. By contrast, our treatment is based on the idea that a suitable resummed version 
of the PN-expanded dynamics, namely the EOB-plus-Pad\'e approach, can, in most cases, 
give an analytical handle on the computation of the inspiral signal down to the 
spin-modified LSSO (and even during the subsequent plunge, as discussed for the spinless 
case in \cite{BD00}). We have argued in several ways that the simple EOB 
Hamiltonian (\ref{eq2.44}) gives reliable 
answers in most cases, and allows one to analytically control the possible amplification 
(or deamplification, when $\widehat{a}_p < 0$) in GW energy loss due to spin effects. 
Moreover, it is only in rather extreme cases that we could agree with \cite{FH98} in 
predicting $\gaq \, 10\% \, M$ energy losses. In most other cases, we think that the EOB 
method provides a reliable basis for computing families of wave forms that will be useful 
templates for the detection of GW's. Another difference with \cite{FH98} is that we have 
argued, on the basis of definite computations, that the spin of the final hole will never 
become nearly extremal (even if the initial spins are extremal). This is important for 
the data analysis of the ring-down signal, because the decay time of the least damped 
quasi-normal-mode starts becoming large only for near extremal holes.

Let us emphasize that the present work is only a first step toward an improved analytical 
understanding of the last stages of inspiral motion of two spinning compact 
object\footnote{Though, in most of the paper we only spoke of binary black holes, it 
should be clarified that our EOB Hamiltonian also applies to binary spinning neutron 
stars or to spinning neutron-star-black-hole systems, at least down to the stage where 
the quadrupole deformation of the neutron star becomes significant.}.
The explicit spin-dependent Hamiltonian (\ref{eq2.44}) (or, better,
$H'_{\rm real}$ defined at the end of Section II)
 has only taken into account the leading effects (in a PN expansion) of the
 spin-dependent interactions. More  work is needed to analytically
 determine  more accurate versions of the EOB Hamiltonian. In particular, it would be
 interesting to explicitly derive the next-to-leading ( 1~PN) corrections to the
 EOB spin-orbit (and spin-spin) interactions.
 Furthermore, we have only provided 
a resummation of the conservative part of the dynamics. There remains the important 
complementary task of resumming the radiation reaction part. This was done in 
\cite{BD00}, using previous results of \cite{DIS98}, only for the spinless case.

Once this is done, we expect, as in our previous study \cite{BD00}, that the presence of 
a LSSO along the sequence of adiabatic orbits will be blurred and will be replaced by a 
continuous transition between inspiral and plunge. There remains also the task of 
studying the effects of spin-dependent interactions on the gravitational waveform emitted 
during the last stages of inspiral and during the plunge (that we have not considered 
here). In other words, one needs to redo, by combining the 
EOB approach with resummed versions of radiation reaction, the studies, valid far from 
the LSO, which were based on straightforward PN-expanded results \cite{kidder}, 
\cite{apostolatos}. Note that our result above about the primary importance of the single 
parameter $\widehat{a}_p$, combined with the understanding \cite{DIS00} that the number of 
``useful'' cycles in the GW signal for massive binaries is rather small, suggests that a 
rather small number of ``spinning templates'' will be really needed in a matched filter 
data analysis. On the other hand, we recall that it was found in \cite{DIS00} that the 
plunge signal (but not the ring-down one, for stellar mass holes) plays a significant role 
in the data analysis.

It will also be interesting to see, within the EOB approach, the extent to which the 
non-linear spin-dependent interactions might, as has been recently suggested 
\cite{levine}, lead to a chaotic dynamical evolution. We a priori suspect that two 
factors will diminish the significance of such chaotic evolutions: (1) they occur only in 
an improbably small region of phase space (involving, in particular, large spins), and 
(2) their effect on the crucial GW phasing is rather small.

It would be very useful to have independent means of testing the accuracy of the EOB 
approach. At this stage we see only three ways of doing that (beyond the performance of 
more internal checks of the robustness of the approach): (i) an analytical calculation of 
the 4~PN interaction Hamiltonian, (ii) a comparison between numerical computations and 
the EOB results, and/or (iii) a comparison between the EOB predictions and the 
forthcoming GW observations. (i) would be important for assessing the convergence of the 
PN-resummed EOB Hamiltonian. In view of the extreme difficulties involved in the 3~PN 
calculations \cite{JS98,JS99,DJS1,DJS3,BF1,BF2,BF3} it would seem hopeless to even 
mention the 4~PN level. But in fact, the EOB approach itself suggests that the current 
methods used in PN calculations are highly inefficient, and unnecessarily complicated. 
Indeed, as emphasized in \cite{DJS2} the final, gauge-invariant content of the 3~PN 
result is contained in only three quantities $a_4$, $b_3$ and $z_3$, and only one of 
them, $a_4 (\nu)$, is really important for determining the dynamics of inspiralling 
quasi-circular binaries. If one could invent a new approximation scheme which computes 
directly $a_4$ (at 3~PN), it might be possible to compute its 4~PN counterpart, $a_5 
(\nu)$. (ii) is not yet possible (at least as a test of the EOB Hamiltonian)
because numerical computations use as initial data 
geometrical configurations that do not take into account most of the crucial physics 
incorporated in PN calculations. Current numerical computations use somewhat ad hoc 
``binary-black-hole-like'' data, often of the restricted spatially conformally flat type, 
without trying to match their initial data to the near LSO configurations predicted by 
(resummed) analytical approaches. On the other hand, let us stress that the value of the 
radial PN-expanded potential $A(\widehat{u}) = 1 - 2 \, \widehat{u} + a_3 (\nu) \, 
\widehat{u}^3 + a_4 (\nu) \, \widehat{u}^4 + 
\cdots$ crucially depends on the non-linear gravitational interactions linked to the 
$h_{ij}^{TT}$ part of the spatial metric, i.e. to its non conformally flat part, and
also to the non-linear interactions linked to the ${\pi}^{ij}_{TT}$ part of the 
gravitational field momenta. For 
instance already at the 2~PN level, the truncation of the Einsteinian prediction for the 
two-body problem (driven into a close orbit by a long past interaction involving retarded 
GW interactions) corresponding to artificially assuming a conformally flat spatial metric 
changes the physically correct value $a_3^{2 \, {\rm PN}} (\nu) = 2 \nu$ into $a^{\rm 
conf. \, flat}_3 (\nu) = \frac{1}{4} \, (18 - 5 \nu) \, \nu$ \cite{DJS2}. [It also
slightly changes the energy map $f$.] For equal-mass 
systems, this corresponds to multiplying the positive $a_3 (\nu)$ by a factor $+ 
2.09375$. As we discussed above, this (artificial) increase of the ``repulsive'' 
character of the non-linear gravitational interactions tends to artificially increase the 
binding of the LSO. As the 2PN coefficient is anyway too small to have a large
impact on the LSO characteristics, this 2PN change does not, by itself,
change much the LSO energy \cite{DJS2}. However, if this tendency to increase the 
``repulsive'' character of the PN expansion (caused by the neglect of the
$h_{ij}^{TT}$-, and ${\pi}^{ij}_{TT}$-dependent interactions) persists at the (numerically
more important) 3PN level, this might explain the current discrepancy between
analytical and numerical estimates of LSO characteristics. 
 In fact, we note that the initial data taken by a recent attempt 
\cite{baker} at fulfilling the proposal of \cite{BD99} to start a full numerical 
calculation only at the moment where it is really needed, i.e. after crossing the LSO, 
uses LSO initial data \cite{baumgarte} with a binding energy $e_{\rm LSO} = E / (m_1 + 
m_2) - 1 \simeq - 
2.3\%$ which is 38\% larger than the value $e_{\rm LSO}^{3 \, {\rm PN}} \simeq - 
1.67\%$ obtained at 3~PN (with $\omega_s = 0$) by analytical estimates. Similarly, the 
LSO orbital period of the initial data of \cite{baker} is $T_{\rm LSO} \simeq 35 \, (m_1 + 
m_2)$ \cite{baumgarte}, which is twice smaller than the 3~PN 
estimate $T_{\rm LSO}^{3 \, {\rm PN}} \simeq 71.2 \, (m_1 + m_2)$ \cite{DJS2}! These 
discrepancies between state-of-the-art numerical LSO initial data and state-of-the-art 
analytical estimates of LSO data are significantly larger than the 
natural ``theoretical error bar'' on the (resummed) 
analytical estimates (derived, say, by comparing 2PN estimates to 3PN ones).
[See, however, the new numerical approach of \cite{GGB} whose LSO data agree well
with the 3~PN EOB estimates \cite{DGG}.]
 In our opinion, this makes it urgent for the numerical relativity community 
to develop ways of constructing initial data that correctly incorporates the crucial 
non-linear physics (linked to the $h_{ij}^{TT}$ and
${\pi}^{ij}_{TT}$ parts of the metric) which is taken into 
account in PN calculations. If a significant discrepancy remains after this is done, one 
will be 
entitled to blame the lack of convergence of the EOB-resummed PN calculations. If one 
finds agreement, this will be a confirmation of the claim made here that 
the Pad\'e-improved EOB is a reliable description of the last orbits before coalescence.
Once one succeeds in matching analytical and numerical results for non-spinning
black holes, it will be very interesting to use numerical data on fast-spinning
 black holes to refine the EOB Hamiltonian by fitting the values of the extra parameters
which can be introduced in the EOB Hamiltonian to represent 
higher PN effects. In the long term
 we expect that such a complementarity between numerical and analytical tools
 will be needed for defining a sufficiently dense set of GW templates. [In view of
 the large dimensionality of the parameter space of the two spinning hole system,
 it seems hopeless to use only numerical techniques to define a dense
 network of templates.]

Finally, even if no decisive progress is made on (i) or (ii) before the first sources are 
detected, there remains the possibility that the first observations might confirm the 
soundness of (or suggest specific modifications of) the EOB-based waveforms, and thereby 
facilitate further detections by narrowing the bank of templates. For instance,
one might include a 4PN contribution $ + a_5 (\nu) u^5$ to $A(u)$, as  
a free parameter in constructing a bank of templates, and 
wait until LIGO/VIRGO/GEO get high signal-to-noise-ratio observations of massive 
coalescing binaries to determine its numerical value.

\acknowledgments
It is a pleasure to thank David Blair for a question which prompted this work, and the 
University of Western Australia for hospitality during its inception.

\end{document}